\documentclass[10pt,preprint]{aastex}
\usepackage{amsmath}
\usepackage{epsfig}

\def\simless{\mathbin{\lower 3pt\hbox
	{$\,\rlap{\raise 5pt\hbox{$\char'074$}}\mathchar"7218\,$}}} 
\def\simgreat{\mathbin{\lower 3pt\hbox
	{$\,\rlap{\raise 5pt\hbox{$\char'076$}}\mathchar"7218\,$}}} 

\newcommand{\band}[2]{\ensuremath{^{{#1}}\!{#2}}}
\newcommand{\lambdaobs}{\lambda_o}
\newcommand{\lambdaemit}{\lambda_e}

\def\beq#1\eeq{\begin{equation}#1\end{equation}}
\def\beqa#1\eeqa{\begin{align}#1\end{align}}
\def\beqq#1\eeqq{\begin{equation*}#1\end{equation*}}
\def\beqqa#1\eeqqa{\begin{align*}#1\end{align*}}
\def\bseq#1\eseq{\begin{subequations}#1\end{subequations}}

\newcommand{\T}{^{\scriptscriptstyle \top}}

\newcommand{\inv}{^{-1}}

\newcommand{\XX}{X}

\newcommand{\Xkn}{X_{kn}}

\renewcommand{\AA}{A}

\newcommand{\WW}{W}

\newcommand{\Wki}{W_{ki}}
\newcommand{\Wkl}{W_{k\ell}}
\newcommand{\HH}{H}

\newcommand{\Hin}{H_{in}}
\newcommand{\Hln}{H_{\ell{}n}}
\newcommand{\BB}{B}

\newcommand{\MM}{M}

\newcommand{\Aki}{A_{ki}}
\newcommand{\Akl}{A_{k\ell}}
\newcommand{\Bij}{B_{ij}}
\newcommand{\Bls}{B_{\ell{}s}}
\newcommand{\Mjn}{M_{jn}}
\newcommand{\Msn}{M_{sn}}
\newcommand{\skn}{\sigma_{kn}}
\newcommand{\PP}{P}
\newcommand{\Pni}{P_{ni}}
\newcommand{\ZZ}{Z}
\newcommand{\Zki}{Z_{ki}}
\newcommand{\Zkl}{Z_{k\ell}}

\newcommand{\Zkj}{Z_{kj}}
\newcommand{\Zks}{Z_{ks}}
\newcommand{\zz}{z}
\newcommand{\yy}{y}
\newcommand{\yn}{y_n}
\newcommand{\yi}{y_i}
\newcommand{\zn}{z_n}
\newcommand{\zi}{z_i}
\newcommand{\betaij}{\beta_{ij}}
\newcommand{\betalj}{\beta_{\ell{}j}}

\newcounter{thefigs}
\newcommand{\fignum}{\arabic{thefigs}}

\newcounter{thetabs}

\newcounter{address}

\slugcomment{To be submitted to \aj}

\shortauthors{Blanton {\it et al.} (2000)}
\shorttitle{$K$-corrections and filter transformations}

\begin{document}
 
\title{$K$-corrections and filter transformations \\
in the ultraviolet, optical, and near infrared}

\author{
Michael R. Blanton\altaffilmark{\ref{NYU}} and 
Sam Roweis\altaffilmark{\ref{UToronto}}
}

\setcounter{address}{0}
\altaffiltext{\theaddress}{
\stepcounter{address}
New York University, Center for Cosmology and Particle Physics, 4 Washington Place, New
York, NY 10003
\label{NYU}}
\setcounter{address}{1}
\altaffiltext{\theaddress}{
\stepcounter{address}
University of Toronto, Dept. of Computer Science, 6 King's College
Rd., Toronto, Ontario M5S 3G4, CANADA
\label{UToronto}}

\clearpage

\begin{abstract}
Template fits to observed galaxy fluxes allow calculation of
$K$-corrections and conversions among observations of galaxies at
various wavelengths. We present a method for creating model-based
template sets given a set of heterogeneous photometric and
spectroscopic galaxy data. Our technique, non-negative matrix
factorization, is akin to principle component analysis (PCA), except
that it is constrained to produce nonnegative templates, it can use a
basis set of models (rather than the delta function basis of PCA), and
it naturally handles uncertainties, missing data, and heterogeneous
data (including broad-band fluxes at various redshifts).  The
particular implementation we present here is suitable for ultraviolet,
optical, and near-infrared observations in the redshift range $0 < z <
1.5$. Since we base our templates on stellar population synthesis
models, the results are intepretable in terms of approximate stellar
masses and star-formation histories. We present templates fit with
this method to data from GALEX, Sloan Digital Sky Survey spectroscopy
and photometry, the Two-Micron All Sky Survey, the Deep Extragalactic
Evolutionary Probe and the Great Observatories Origins Deep Survey.
In addition, we present software for using such data to estimate
$K$-corrections and stellar masses.
\end{abstract}


\section{Motivation}
\label{motivation}

New surveys at low and high redshift have provided us with estimates
of galaxy spectral energy distributions (SEDs) for an enormous number
of galaxies. When comparing populations of galaxies at different
redshifts in these surveys, we need to use comparable measurements of
the galaxy SEDs. However, different surveys use different bandpasses
and the restframe wavelengths of these bandpasses necessarily vary
with redshift. We need to be able to handle this heterogeneity in
order to make sensible comparisons among all of these new surveys.

In this paper, we present a method for doing so, by calculating
$K$-corrections between observed and desired bandpasses.  The
$K$-correction between a bandpass $R$ used to observe a galaxy at
redshift $z$ and the desired bandpass $Q$ is defined by the equation
(\citealt{oke68a, hogg02a}):
\begin{equation}
m_R = M_Q + \mathrm{DM}(z) + K_{QR}(z) - 5 \log_{10} h 
\end{equation}
where $\mathrm{DM}(z) = 25 - 5\log_{10} (d_L / (h^{-1}{\mathrm{~Mpc}}))$ is the
bolometric distance modulus calculated from the luminosity distance
$d_L$, and $M_Q$ is the absolute magnitude. The absolute magnitude is
defined as the apparent magnitude an object would have if were
observed 10 pc away, in bandpass $Q$, at rest.  The traditional
definition of the $K$-correction takes $Q=R$. However, we note that in
practice many surveys do perform $K$-corrections from one observed
bandpass $R$ to another bandpass $Q$ in the rest frame. This practice
is particularly common when dealing with high redshift observations.
In addition to $K$-corrections, this method also provides an
interpretation of the data in terms of a physical model which
describes the stellar mass and star-formation history of each galaxy.

The method is designed to work well for a wide range of data sets.  It
uses photometry and spectroscopy of the Sloan Digital Sky Survey
(SDSS; \citealt{york00a}) the Galaxy Evolution Explorer (GALEX;
\citealt{martin05a}) in the ultraviolet, and the Two-Micron All Sky
Survey (2MASS; \citealt{skrutskie97a}) in the near infrared (NIR). In
addition, at higher redshifts, we use constraints from the Deep
Extragalactic Evolutionary Probe 2 (DEEP2; \citealt{davis03a,
faber03a}) and the Great Observatories Origins Deep Survey (GOODS;
\citealt{giavalisco04a}).  These and other data sets provide a huge
set of information about galaxy colors and spectra which we can use to
help understand their star-formation histories.

We note that some of the algorithmic techniques used here may have
applications in other areas of astrophysics. First, nonnegative matrix
factorization (NMF; \citealt{lee00a}), and the extensions to it we
describe here, is a particularly useful variant on principal component
analysis. Second, the nonnegative least squares algorithm of
\citet{sha02a} has the virtue of being extremely simple to implement.
These methods may find other applications in image analysis, spectroscopic
analysis, and model fitting in astrophysics.

We also release the templates in electronic form as well as an
implementation of the methods used to fit the templates to data. This
software {\tt kcorrect v4\_1} is distributed on the World Wide
Web{\footnote {\tt http://cosmo.nyu.edu/blanton/kcorrect/}}. It
consists of a core C library which performs most of the complex and
computationally intensive tasks, plus an IDL library that provides a
high-level interface.  This software is an update of two major earlier
releases ({\tt v1\_16}; \citealt{blanton03b}; {\tt v3\_2}). The
improvement over the previous version is twofold. First, we have
improved the templates such that they successfully fit galaxies in the
restframe UV, as observed by GALEX, DEEP2, and GOODS. Second, because
the templates are now completely model-based, the fits have a physical
interpretation in terms of a star-formation history. The IDL library
also has a number of new useful functions.

In Section \ref{algorithm}, we describe how to find suitable
templates, given a set of models and a set of data. We also describe
the data and models used here. In Section \ref{results}, we show the
results: the best-fit templates and how well they fit the data.  In
Section \ref{kcorrect}, we describe how to convert our results into an
estimate of the $K$-correction. In Section \ref{physical}, we describe
the physical interpretation of the templates and of fits to data.  In
Section \ref{sdss2bessell}, we present some simplified, linear
transformations between various bandpasses. In Section \ref{summary},
we summarize our results. Appendix \ref{nmf} describes the NMF
algorithm.  Appendix \ref{format} describes the electronic format of
our results.

Where necessary, we have assumed cosmological parameters $\Omega_0 =
0.3$, $\Omega_\Lambda = 0.7$, and $H_0 = 100$ $h$ km s$^{-1}$
Mpc$^{-1}$ (with $h=1$), unless otherwise noted. All magnitudes are
(unless otherwise noted) AB-relative.  Except where noted, I will
always be referring to the version of the {\tt kcorrect} software
labeled {\tt v4\_1\_4}.

\section{Finding the templates}
\label{algorithm}

\subsection{Overview}

From the spectroscopic observations we know that galaxy spectra reside
in a low-dimensional subspace. Principal Component Analysis (PCA) of
galaxy spectra, introduced by \citet{connolly95b} and applied many
times since then, demonstrates that most of the variance in the
distribution of galaxies in spectral space can be explained using a
few templates. This means that, in the linear space of all possible
spectra, galaxies exist in only a small subspace. Therefore, even with
very heterogeneous data, we should be able to determine the properties
of this subspace.

Here we present an approach to combining heterogeneous data in order
to determine the properties of the subspace of galaxy spectra. Rather
than taking the model-free approach used by PCA, we here restrict the
space of possible spectra to those predicted from the high resolution
stellar population synthesis model of \citet{bruzual03a} and the
nebular emission line models of \citet{kewley01a}.  This approach both
constrains the problem appropriately and yields a natural theoretical
interpretation of the results in terms of star-formation histories.

In a nutshell, our algorithm does the following. Given the
observations (and uncertainties of those observations) available for
each galaxy, it finds the nonnegative linear combination of $N$
template star-formation histories which best predict those
observations in the $\chi^2$ sense. Given the entire set of galaxy
observations available, it also fits for the $N$ template
star-formation histories. The technical name for this algorithm is
Nonnegative Matrix Factorization (NMF; \citealt{lee99a, lee00a}), and
we describe it in detail in Appendix \ref{nmf}. Note that this problem
is not the same as the well-known nonnegative least squares problem,
which would fit for the best-fit nonnegative linear combination of
templates, but not for the form of the templates themselves.

This approach is similar to PCA in that it finds the small spectral
subspace in which galaxies exist, and can in some ways be thought of
as ``nonnegative PCA.''  However, our method has several advantages
over the standard PCA approach. First, the results yield, along with a
subspace in the space of all possible spectra, a natural
interpretation, which is the corresponding subspace in the space of
all possible star-formation histories. Second, it naturally handles
data uncertainties and missing data, which allows it to ignore that
variation which is due purely to statistical errors.  Third, it
handles the complications of observing galaxy spectra photometrically
using broad-band filters of galaxies at varying redshifts, whereas
standard PCA requires a constant grid of observations in rest-frame
wavelength space.

In the following subsections, we introduce the basis set of models we
use here (\S\ref{models}), the data we fit to (\S\ref{data}), and how
we combine the two to set up the NMF problem (\S\ref{setupnmf}).  In
Appendix \ref{nmf} we describe how to actually solve to NMF problem.

\subsection{The models}
\label{models}

We begin with a basis set of 485 spectral templates. Of these, 450 are
a set of instantaneous bursts from \citet{bruzual03a}, using the
\citet{chabrier03a} stellar initial mass function and the Padova1994
isochrones. We select 6 metallicities (0.005, 0.02, 0.2, 0.4, 1, and
2.5 times solar). For each metallicity we select 25 ages (between 1
Myr and 13.75 Gyr, spaced almost logarithmically in age).  For each
age and metallicity we make 3 choices of dust model: (1) no dust
extinction; (2) $\tau_V = 3$ dust with Milky Way type extinction (as
input into the models of
\citealt{witt00a} with a ``homogeneous'' distribution and ``shell''
geometry); (3) $\tau_V = 3$ dust with SMC type extinction (from the
same models). We smooth each template to $300$ km s$^{-1}$ velocity
dispersion (we will smooth the SDSS spectra in the training set to the
same resolution). 

The remaining 35 templates are from MAPPINGS-III (\citealt{kewley01a})
models of emission from ionized gas. We choose the predictions for an
8 Myr old continuous star-formation history with 5 possible
metalliciticites (0.5, 0.2, 0.4, 1.0, and 2. times solar) and 7
possible ionization parameters ($q= 5\times 10^6$, $10^7$, $2\times
10^7$, $4\times 10^7$, $8 \times 10^7$, $1.5\times 10^8$, and $3\times
10^8$ cm s$^{-1}$). We take the spectra (given as a set of delta
functions) and smooth them to $300$ km s$^{-1}$ velocity
dispersion. The one alteration we make to all of these templates is to
remove Ly$\alpha$, because it is generally much larger than observed
in real galaxies.

Let us refer to these 485 basis templates as $M_{j}(\lambda)$,
expressed in units of ergs s$^{-1}$ \AA$^{-1}$.

We seek to reduce this full basis space to a subspace in which
galaxies actually exist. In particular, we seek five templates
$F_{i}(\lambda)$ built from nonnegative combinations of the
original basis set of $N$ templates:
\begin{equation}
F_{i}(\lambda) = \sum_j b_{ij} M_{j}(\lambda),
\end{equation}
in units of ergs s$^{-1}$ \AA$^{-1}$. In principle, we could seek any
number of templates; from our experiments, we have found that five
turns out to be a number large enough to explain the data we use
here. For each galaxy $k$ we want our model
${\hat{F}_\lambda}(\lambda)$ for their spectrum to be a nonegative sum
of these five templates:
\begin{equation}
{\hat{F}}_{k}(\lambda) = \sum_i a_{ki} F_{i}(\lambda).
\end{equation}

\subsection{The data}
\label{data}

The training set consists of:
\begin{enumerate}
\item SDSS spectroscopic data in the observed range $3800 < \lambda <
9000$ \AA, for 400 Luminous Red Galaxies between $0.15 < z < 0.5$ (LRGs;
\citealt{eisenstein01a}) and 1600 Main sample galaxies between $0.001
< z < 0.4$ (\citealt{strauss02a})
\item SDSS photometric data on an independent set of LRGs ($griz$
photometry only) and Main sample galaxies (using the full $ugriz$
photometry) in the same redshift ranges. We use 2000 LRGs and 7000
Main sample galaxies.  For these galaxies we include 2MASS $JHK_s$
extended source catalog data (\citealt{jarrett00a}) where available.
\item GALEX DR1 far UV ($\sim 1500$ \AA) and near UV ($\sim 2300$ \AA)
photometry for Main sample SDSS galaxies with redshifts and $ugriz$
photometry (4000 galaxies; \citealt{martin05a}).
\item The $BRI$ photometry for high redshift galaxies in the DEEP2 
DR1 release between $0.6 < z < 1.5$ (4000
galaxies; \citealt{davis03a, faber03a}).
\item The $BVizJHK_s$ photometry for GOODS galaxies between $0.5 < z < 2$ (1000
galaxies; \citealt{giavalisco04a}).
\end{enumerate}
The SDSS, 2MASS, and GALEX galaxies were selected from and the matches
were obtained by the New York University Value-Added Galaxy Catalog
(NYU-VAGC;
\citealt{blanton05a}).  

The SDSS data processing consists of astrometry \citep{pier03a};
source identification, deblending and photometry \citep{lupton01a};
photometricity determination \citep{hogg01a}; calibration
\citep{fukugita96a,smith02a}; spectroscopic target selection
\citep{eisenstein01a,strauss02a,richards02a}; spectroscopic fiber
placement \citep{blanton03a}; and spectroscopic data reduction.  We
recalibrated our photometry using the ``ubercalibration'' procedure
described in \citet{blanton05a}.  Descriptions of these pipelines also
exist in \citet{stoughton02a}.  An automated pipeline called {\tt
idlspec2d} (in this case {\tt v4\_9}) measures the redshifts and
classifies the reduced spectra (Schlegel et al., in preparation). We
use the Petrosian magnitudes for the SDSS data (except where noted
below for the LRG-only sample).

For the GALEX and GOODS data we use ``auto'' magnitudes, which are the
Kron-like magnitudes from SExtractor (\citealt{bertin96a}).  For 2MASS
we use the ``extrapolated'' magnitudes from the Extended Source
Catalog (\citealt{jarrett00a}).

Note that there are 20 different broadband photometric filters listed
above ($B$ in DEEP2 is different than the $B$ used by GOODS). The {\tt
kcorrect} product described in Section \ref{summary} contains a
tabulation of the response functions for all of these filters.

\subsection{Comparing data and models}
\label{setupnmf}

The data consist of spectra and bandband photometric measurements of
galaxies at a number of redshifts, and we have to relate the models to
these measurements.

For the spectra, we take the observed spectrum
$f_{k}(\lambda)$ (in ergs cm$^{-2}$ s$^{-1}$ \AA$^{-1}$) of
each galaxy $k$ at redshift $z$ (corresponding to a luminosity
distance $d_L(z)$ in cm; see \citealt{hogg99a}) and calculate the
restframe luminosity per unit wavelength:
\begin{equation}
{{F}}_{k}(\lambda) = 
{{f}}_{k}[\lambda (1+z)](1+z) (4 \pi d_L^2),
\end{equation}
That is, the spectrum is shifted due to the redshift, while the
integral of the numerator over wavelength (the total luminosity) is
constant with redshift, and the total flux is related to the total
luminosity by the inverse square law.  In addition, we smooth each
spectrum by such an amount that, given its estimated velocity
dispersion, its total velocity dispersion after smoothing is 300 km
s$^{-1}$ (there are a small number of galaxies with larger velocity
dispersions, which we leave unchanged).

Note that if we discretize the spectra to wavelengths $\lambda_l$, the
relationship between the predicted SED for a galaxy $k$ and the basis
set $M_{jl}$, becomes simply:
\begin{equation}
\hat{F}_{kl} = \sum_{ij} a_{ki} b_{ij} M_{jl}
\end{equation}

If one has a spectrum for a galaxy the
expression for the contribution to $\chi^2$ from each wavelength
$\lambda_l$ in the spectrum is then quite simple:
\begin{eqnarray}
\label{chi2spec}
\chi_{kl}^2 &=& \left[\frac{F_{k}(\lambda_l) -
{\hat{F}}_{k}(\lambda_l)} 
{\sigma^2_k(\lambda_l)}\right]^2 \cr
&=&
\left[\frac{F_{k}(\lambda_l) -
\sum_{ij} a_{ki} b_{ij} M_{jl}
}{\sigma^2_k(\lambda_l)}\right]^2. 
\end{eqnarray}

Comparing observed broadband flux measurements is a bit more
complicated, because it is a projection of the spectrum onto the
broadband filter $p$ at the observed redshift $z$. Instead of
adjusting the observed fluxes as we could so easily do for the spectra
above, for the photometry we express the models in terms of predicted
broadband fluxes at each redshift.

Here we express the flux for each galaxy $k$ in units of AB maggies
$\mu_p$, which are defined as what one would measure in bandpass $p$
relative to the AB standard source. For example, if we transform our
spectral energy density basis function $M_{j}(\lambda)$ to a flux
density $m_{j}(\lambda)$, we can calculate the contribution of
that basis function to the predicted maggies as:
\begin{equation}
\label{maggies}
\mu_{jp} = 
\frac{\int_0^{\infty} d\lambda \lambda R_p(\lambda) {m}_{j}(\lambda)}
{\int_0^\infty d\lambda \lambda R_p(\lambda) f_{\mathrm{AB}}(\lambda) }
\end{equation}
Here, the response function $R_p(\lambda)$ is proportional to the
contribution to the detector signal of a photon with wavelength
$\lambda$ entering the Earth's atmosphere (or entering the telescope
for a space telescope). The AB standard source is $f_{
\mathrm{AB}}(\lambda) d\lambda = f_{
\mathrm{AB}} (\nu) d\nu$ and $f_{\mathrm{AB}}(\nu)=3631$ Jy $= 3.631 \times
10^{-20}$ erg s$^{-1}$ cm$^{-2}$ Hz$^{-1}$. Of course maggies $\mu$
are related to magnitudes $m$ as:
\begin{equation}
m = -2.5 \log_{10} \mu, 
\end{equation}
such that the AB standard source would (if it existed) have $\mu=1$
and $m=0$ for all bandpasses.

In this context it is worth noting that many authors
(e.g. \citealt{bessell90a}) tabulate the contribution to the detector
signal per {\it unit of energy} in photons of wavelength $\lambda$
instead of per {\it photon} with wavelength $\lambda$.  We will refer here
to the former quantity as $R_p'(\lambda)$, though in the literature it is
often referred to without a prime (and without any explicit
definition!). Clearly $R_p(\lambda) \propto R_p'(\lambda)/\lambda$,
since the higher the frequency of the photon, at a fixed response per
unit energy there is a higher response per photon. With this
substitution, one can reexpress Equation
\ref{maggies} appropriately in terms of $R_p'(\lambda)$.  
Generally, though not universally, authors tabulate $R_p'(\lambda)$
for bandpasses whose standards were originally calibrated using
energy-counting devices rather than photon-counting devices. However,
from the point of view of the analysis of the observations it is
irrelevant what the devices used for the standards and the
observations are, as long as one calculates $R_p(\lambda)$ and uses
Equation \ref{maggies}.  

The prediction for the broadband fluxes from Equation
\ref{maggies} is only for a specific redshift $z$. It turns out to
simplify our mathematics to calculate the projection of each basis
function $j$ onto each filter $p$ for a grid of redshifts (in this
case spaced by 0.005 between redshifts 0 and 2). Thus, below we will
take $p$ to index all the filters at all such redshifts.

Just as before, we can now write down the relationship between the
predicted broadband flux and the basis set $\mu_{jp}$, in the bandpass
and redshift corresponding to the index $p$, for a galaxy $k$:
\begin{equation}
\hat{\mu}_{kp} = \sum_{ij} a_{ki} b_{ij} \eta_{jp}
\end{equation}

The contribution to the total $\chi^2$ of a broadband flux is
therefore just:
\begin{eqnarray}
\label{chi2photo}
\chi^2_{kp} &=& \left[\frac{\mu_{kp} - \hat\mu_{kp}}
{\sigma_{kp}}\right]^2 \cr
&=&
\left[\frac{\mu_{kp} -
\sum_{ij} a_{ki} b_{ij} \eta_{jp}
}{\sigma^2_{kp}}\right]^2. 
\end{eqnarray}
For a given galaxy we do not have every filter, and we only have an
observation of each filter at a single redshift. We pick the closest
redshift on the redshift grid, and use the measured filters at that
redshift for our expression of $\chi^2$, and set $1/\sigma_{kp}^2 = 0$
for the rest of the values of the index $p$ so that we zero-weight
those predictions.

The matrices $\eta_{jp}$ (related to the broad-band fluxes) and
$M_{jl}$ (related to the spectra) are totally fixed, and we combine
them into a single matrix $M_{jn}$ and the indices $p$ and $l$ into a
single index $n$, to handle both the broad-band fluxes and spectra
simultaneously. We can similarly combine our observations $\mu_{kp}$
and $F_{k}(\lambda_l)$ and their uncertainties $\sigma_{kp}$
and $\sigma_{k}(\lambda_l)$ into vectors $x_{kn}$ and $\sigma_{kn}$.
Then we can combine the Equations \ref{chi2spec} and
\ref{chi2photo} into a single equation:
\begin{equation}
\label{chi2}
\chi^2 = \sum_{kn} \left[ 
\frac{x_{kn} - \sum_{ij} a_{ki} b_{ij} M_{jn}}
{\sigma_{kn}} \right]^2
\end{equation}
There is a simple method that we describe in Appendix \ref{nmf} called
nonnegative matrix factorization (NMF) to iterate to the nonnegative
$a_{ki}$ and $b_{ij}$ which locally minimize Equation \ref{chi2}. The
basic method is implemented in a public piece of code named {\tt
NMF\_SPARSE} in the {\tt idlutils} distribution of IDL
utilities.\footnote{\tt http://skymaps.info}. As the name implies, our
implementation takes advantage of the fact that many of the matrix
operations are on very sparse matrices (for example, for each galaxy
with photometric data there are {\it no} spectroscopic data points and
{\it only} photometric data at a single redshift).

As we describe in Appendix \ref{nmf}, the NMF problem is not convex:
that is, there are multiple local minima. Our method finds one of the
local minima, but is not guaranteed to find the global minimum. Thus,
the reader who tries to reproduce our results using our methods or
others may (depending on their initial conditions) find different
local minima of $\chi^2$ than do we.

Once we have fit for $b_{ij}$ using the training set, we can minimize
Equation \ref{chi2} for any other galaxy using any nonnegative least
squares algorithm to determine $a_{ki}$ (since the minimization of
Equation \ref{chi2} has a linear form in that case).  When we do so
here we use the beautifully simple iterative method of \citet{sha02a}.

\section{Results}
\label{results}

\subsection{An example: the Luminous Red Galaxy templates}

We begin with the simplest case, which is fitting a {\it single}
template to the photometric data of the LRG sample
(\citealt{eisenstein01a}). For many of these galaxies, which extend to
$r<19.5$ and are intrinsically red, the $u$-band flux is extremely
poorly measured, so we ignore the $u$-band for all LRGs. 

\begin{figure}
\figurenum{\fignum}
\plotone{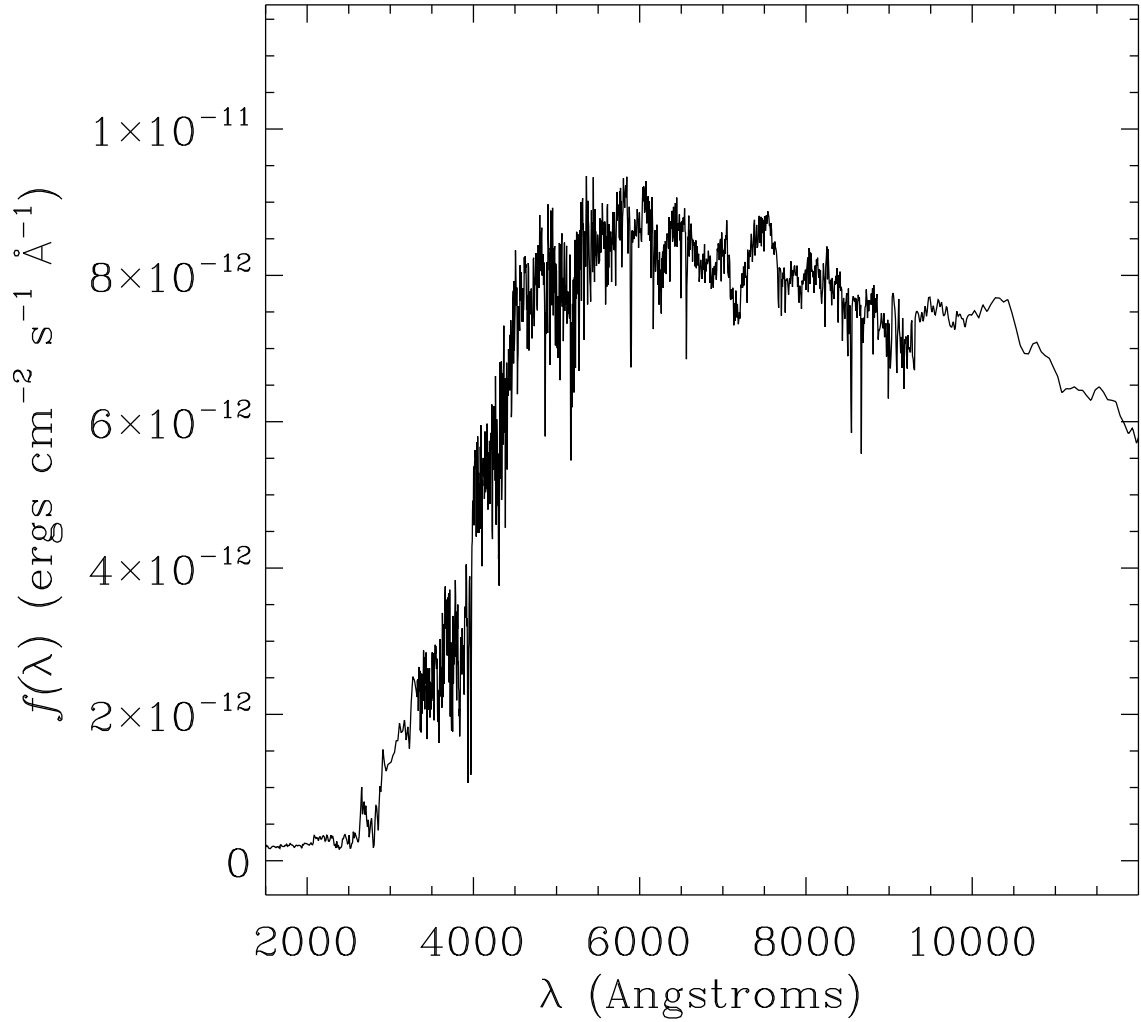}
\caption{\label{spec_lrg} Best fit LRG spectral template in the rest
frame (normalization is for a 1 $M_\odot$ galaxy located 10 pc away,
or equivalently a $10^{12}$ $M_\odot$ galaxy located 10 Mpc away).}
\end{figure}

Figure \ref{spec_lrg} shows the spectrum of the best-fit LRG.  This
template is constrained by data between about 2000 \AA\ and 10000 \AA;
outside that range it is an extrapolation.

\begin{figure}
\figurenum{\fignum}
\plotone{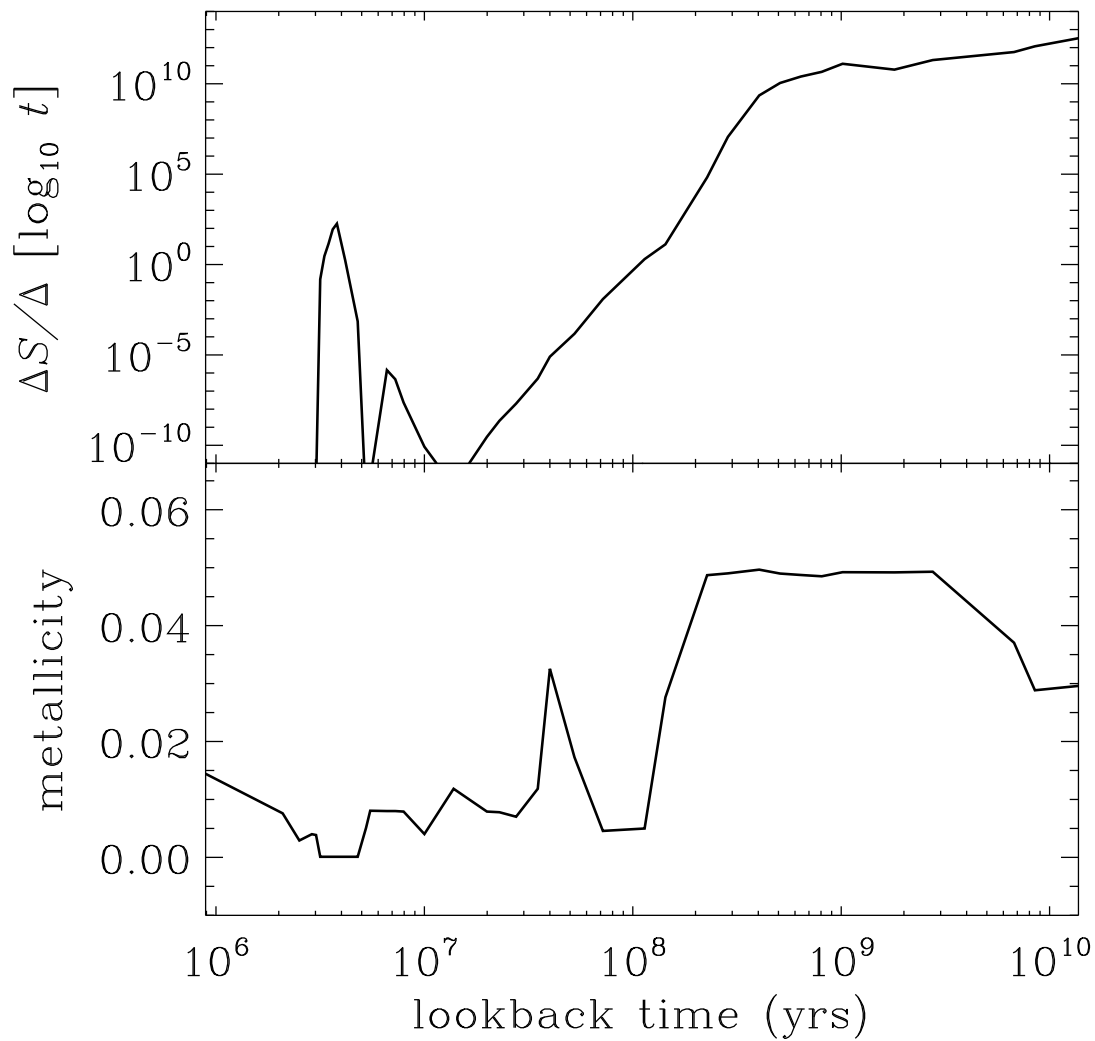}
\caption{\label{sfh_lrg} Star-formation history corresponding to LRG
	spectral template of Figure \ref{spec_lrg}. Top panel shows the
	number of stars formed per logarithmic time interval ($t$ is
	expressed in years, curve is normalized for a $10^{12}$ $M_\odot$
	galaxy). Almost all of the stars are formed in the first couple of
	billion years --- note that the recent ``spike'' is represents a
	tiny fraction ($\sim 10^{-8}$) of the total number of stars. Bottom
	panel shows the mean metallicity of the population as a function of
	time. Note that the details of these functions are rather poorly
	constrained. }
\end{figure}

Figure \ref{sfh_lrg} shows the star-formation history corresponding to
this best-fit LRG spectrum. The top panel shows the star-formation per
unit time as a function of look-back time.  The bottom panel shows the
mean metallicity of the forming stars as a function of time.  When
considering these plots and those below, do remember that although
these results are our best-fit, the parameters are highly degenerate
(especially in, for example, neighboring age bins).  We show these
merely to illustrate the general nature of the fits.

\begin{figure}
\figurenum{\fignum}
\plotone{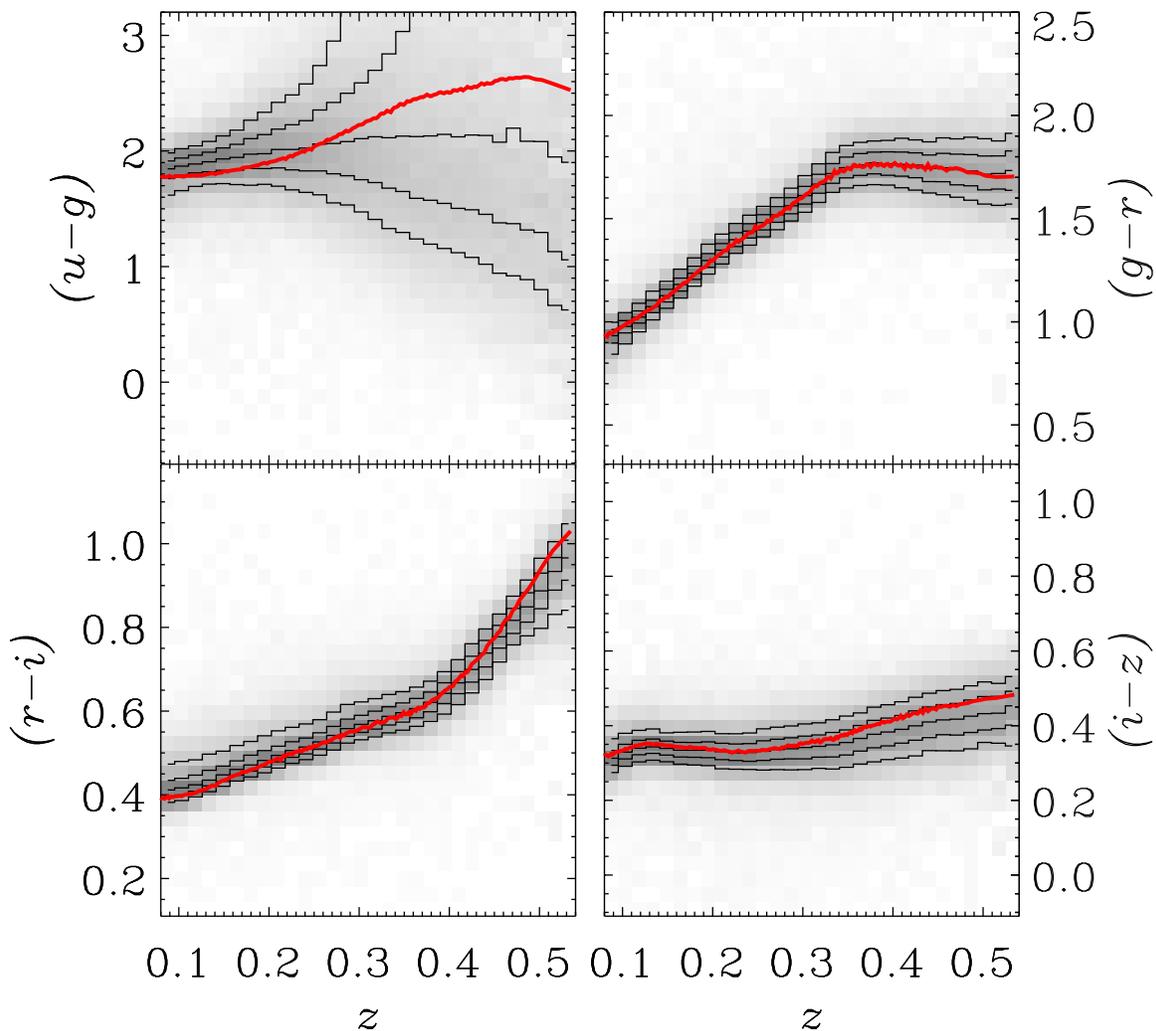}
\caption{\label{lrg_colors} SDSS colors of LRGs as a function of
	redshift. The greyscale is the conditional distribution of color
	within each redshift bin. The thin lines are the 10\%, 25\%, 50\%,
	75\%, and 90\% quantiles of the distribution. The thick line
	is the prediction of the model. The $u$ band is not included in the
	fit, and the $u$ magnitudes of most LRGs are poorly known. The other
	colors fit the models reasonably well. This model, remember, is
	given incredible freedom, meaning that the above agreement is the
	best one can do with the stellar population synthesis code of
	\citet{bruzual03a}. }
\end{figure}

Figure \ref{lrg_colors} shows the LRG colors as a function of
redshift, with the best fit template color overplotted as the smooth,
thick line. In the code described in Section \ref{summary}, we used the
routine {\tt sdss\_kcorrect} (with the optional flag {\tt /lrg}) to
perform these fits. The best fit is a good fit to the data.

Note that in this example we have fit a nonevolving template, which is
inappropriate over this range of redshifts --- even if it is a good
fit to the colors! In principle, we can adjust the methods used here
for the case of evolving templates, but we will not do so here.

In Appendix \ref{format} we describe the form in which we release 
the star-formation histories and spectra associated with this
template.

\subsection{Templates for the full data set}

\begin{figure}
\figurenum{\fignum}
\plotone{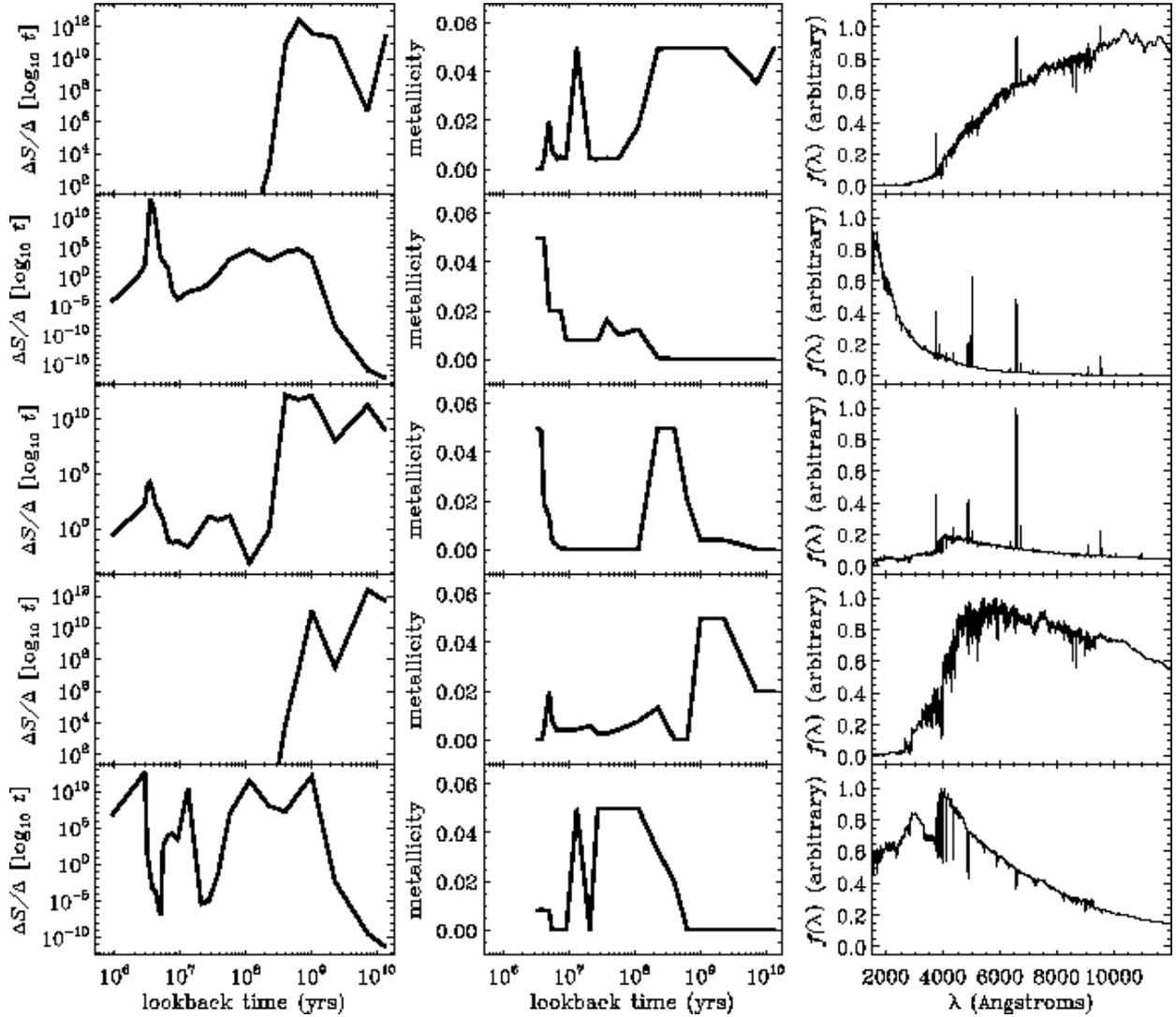}
\caption{\label{sfh_templates} Similar to Figure \ref{sfh_lrg}, but
for the five global templates. Again, there are many degeneracies in
the fit parameters (though there are fewer in the actual spectra), so
these figures need to be interpreted appropriately.}
\end{figure}

We next apply the method to the full data set and fit for five
templates. From experiments, we found that we needed five to
adequately explain the data. More than five templates would allow us
to explain the data better, but only marginally so.  The right-most
column of Figure \ref{sfh_templates} shows the spectrum from the
ultraviolet through the near infrared for the resulting
templates. Note that there is a very old template, a very young
template, and several intermediate templates, including one which is
close to that of an A star.

The left and middle columns of Figure \ref{sfh_templates} show the
star-formation histories and metallicities associated with the
templates. We have not imposed any smoothness criterion on these fits,
which explains the ragged appearance of these histories.  The details
of these fits are weakly constrained due to degeneracies within and
among the templates, but we show them here for completeness.

In all the results below, we use these five templates, unaltered. That
is, when we speak below of ``fitting'' the templates, we mean we fix
$b_{ij}$ to the five templates shown in this section and fit only for
$a_{ki}$. Since we have determined these templates to be appropriate
for the data sets we include in the training set, they end up
providing good fits to most other galaxies from the parent data sets.
In general below, the tests we perform are not to the training set but
to independent sets of galaxies.

In Appendix \ref{format} we describe the form in which we release the
star-formation histories and spectra associated with these templates.

\subsection{Explanatory power of templates}

\begin{figure}
\figurenum{\fignum}
\plotone{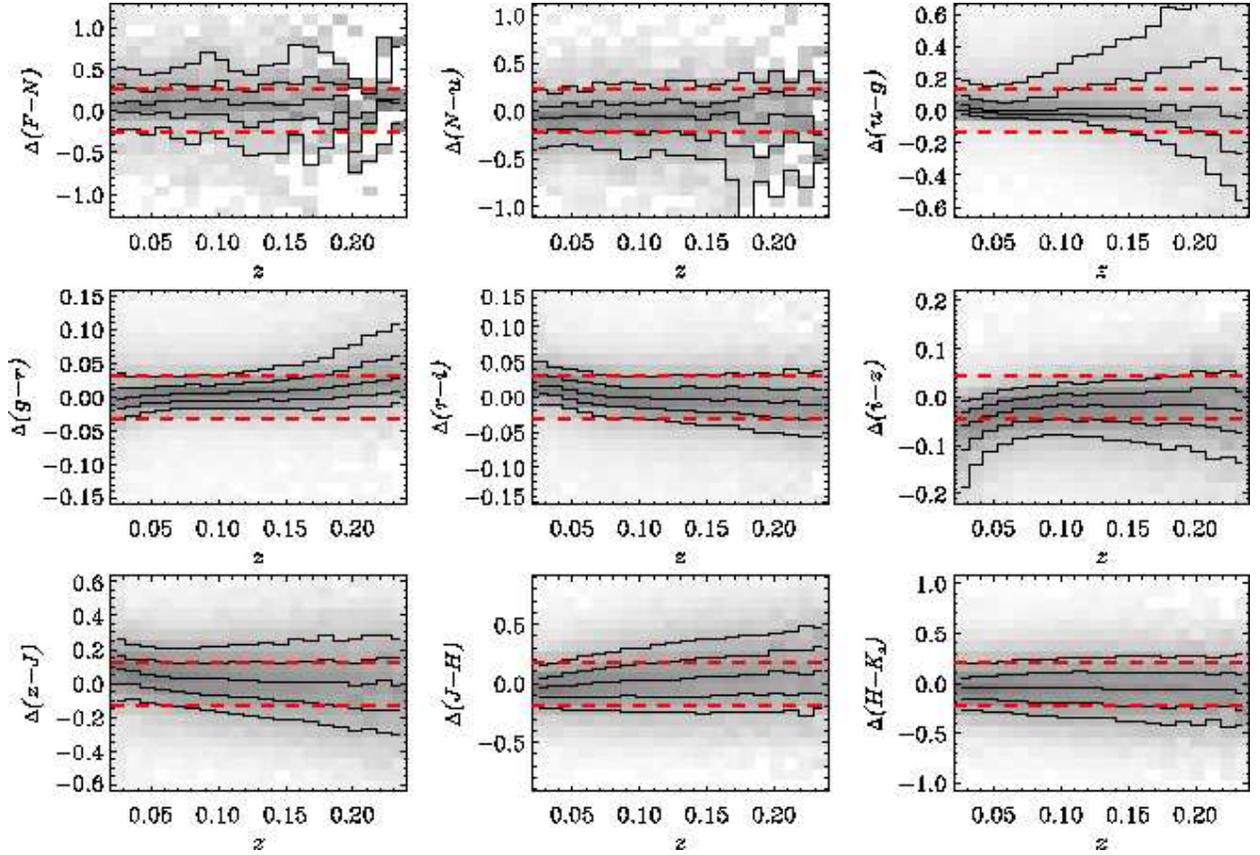}
\caption{\label{fullfits} Color residuals (defined explicitly in the
	text) of GALEX, SDSS, and 2MASS observations relative to our best
	fit 5-template model.  The greyscale is the conditional distribution
	of the color residual given the redshift.  The thin lines are the
	10\%, 25\%, 50\%, 75\%, and 90\% quantiles of the distribution. The
	thick dashed lines show the estimated 1$\sigma$ uncertainties in the
	colors from the photometric catalogs. Relative to the uncertainties,
	there are no significant biases or redshift trends in these fits. }
\end{figure}

These templates explain the photometric data rather well. For example,
consider Figure \ref{fullfits}, showing the color residuals of the
observations with respect to the best fits, when fitting to galaxies
with GALEX, SDSS, and 2MASS data. In this context we define the color
residuals as (for example):
\begin{equation}
\Delta [u-g] = [u_{\mathrm{obs}}- g_{\mathrm{obs}}] -
[u_{\mathrm{model}}- g_{\mathrm{model}}] .
\end{equation}
In the code described in Section \ref{summary}, we used the routine {\tt
gst\_kcorrect} to perform these fits.  In Figure \ref{fullfits} the
thick dashed lines show the estimated 1$\sigma$ uncertainties in the
colors from the photometric catalogs. Relative to the uncertainties,
there are no significant biases or redshift trends in these fits. Note
that the galaxies used in this test are distinct from the training set
on which we fit the five templates.

\begin{figure}
\figurenum{\fignum}
\plotone{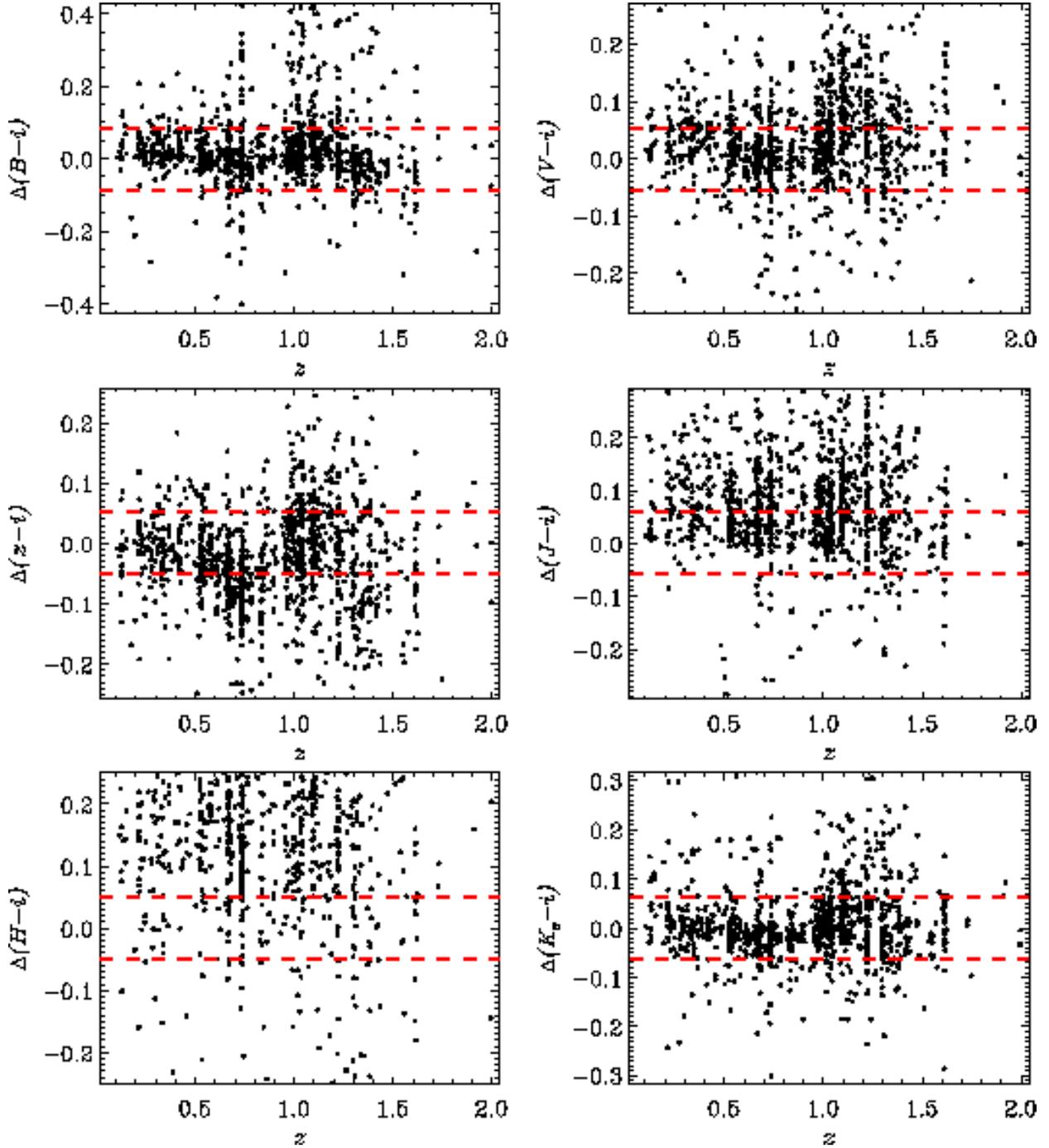}
\caption{\label{goods} Color residuals in fit using the standard five
templates to GOODS data, compared to the typical uncertainties (thick
dashed lines). Note that the fits always do poorly on the $H$ band,
which we believe to be a catalog error. }
\end{figure}

\begin{figure}
\figurenum{\fignum}
\plotone{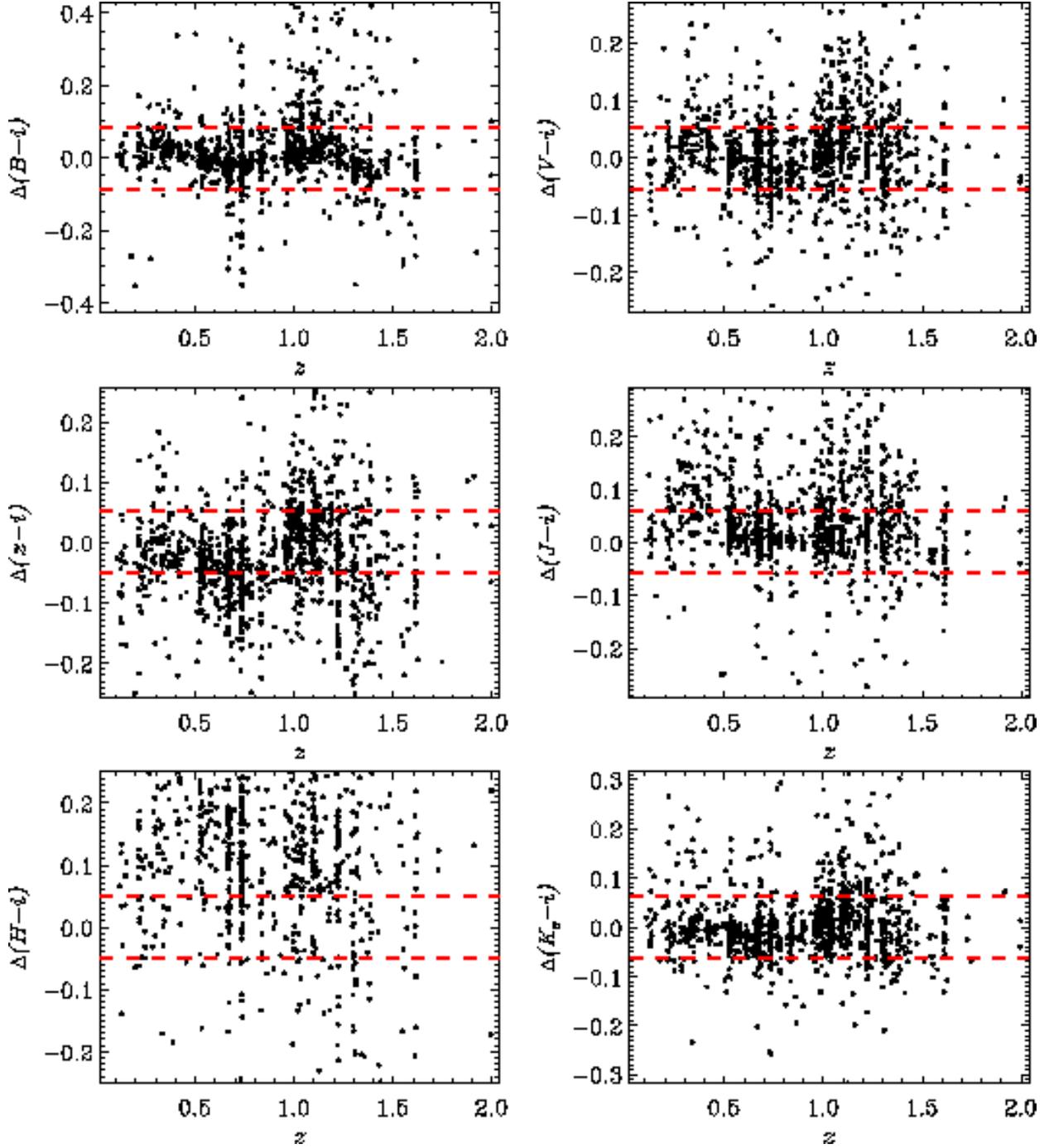}
\caption{\label{goods_special} Same as Figure \ref{goods}, but fitting
using five templates specially designed for GOODS. These templates
have smaller residuals in many respects but still fail to fit the $H$
band data. }
\end{figure}

The templates do well on higher redshift data, as well. For example,
Figure \ref{goods} shows the color residuals for GOODS data. Here we
find that there are somewhat larger residuals with respect to the
uncertainties (again shown as the thick dashed lines). We can compare
this result to that in Figure \ref{goods_special}, which we get by
fitting a new set of templates to {\it only} the GOODS data (this
alternate set of five templates is also distributed in the manner that
Appendix \ref{format} describes). From Figure \ref{goods_special}, we
conclude that {\it some} of the errors are irreducible, and require
either a larger set of templates for high redshift galaxies (a
possibility we find a priori unlikely), or that there are simply
errors in the catalogs or the input filter curves. For example, in the
$z$ and $V$ bands we find significant trends between redshifts $z=0.5$
and $1.5$ (echoed in Figure \ref{goods}). More significantly, in the
$H$ band there is at least a 10\% offset in the magnitudes across all
redshifts, which is almost certainly a catalog or filter curve
error. Other errors (such as a large scatter in the residuals in the
$B$ band and some redshift dependence in the $J$ band) are clearly due
to the fact that the templates are not primarily designed for GOODS
data. All in all, we recommend using the special templates for
$K$-corrections within the GOODS data set.

\begin{figure}
\figurenum{\fignum}
\plotone{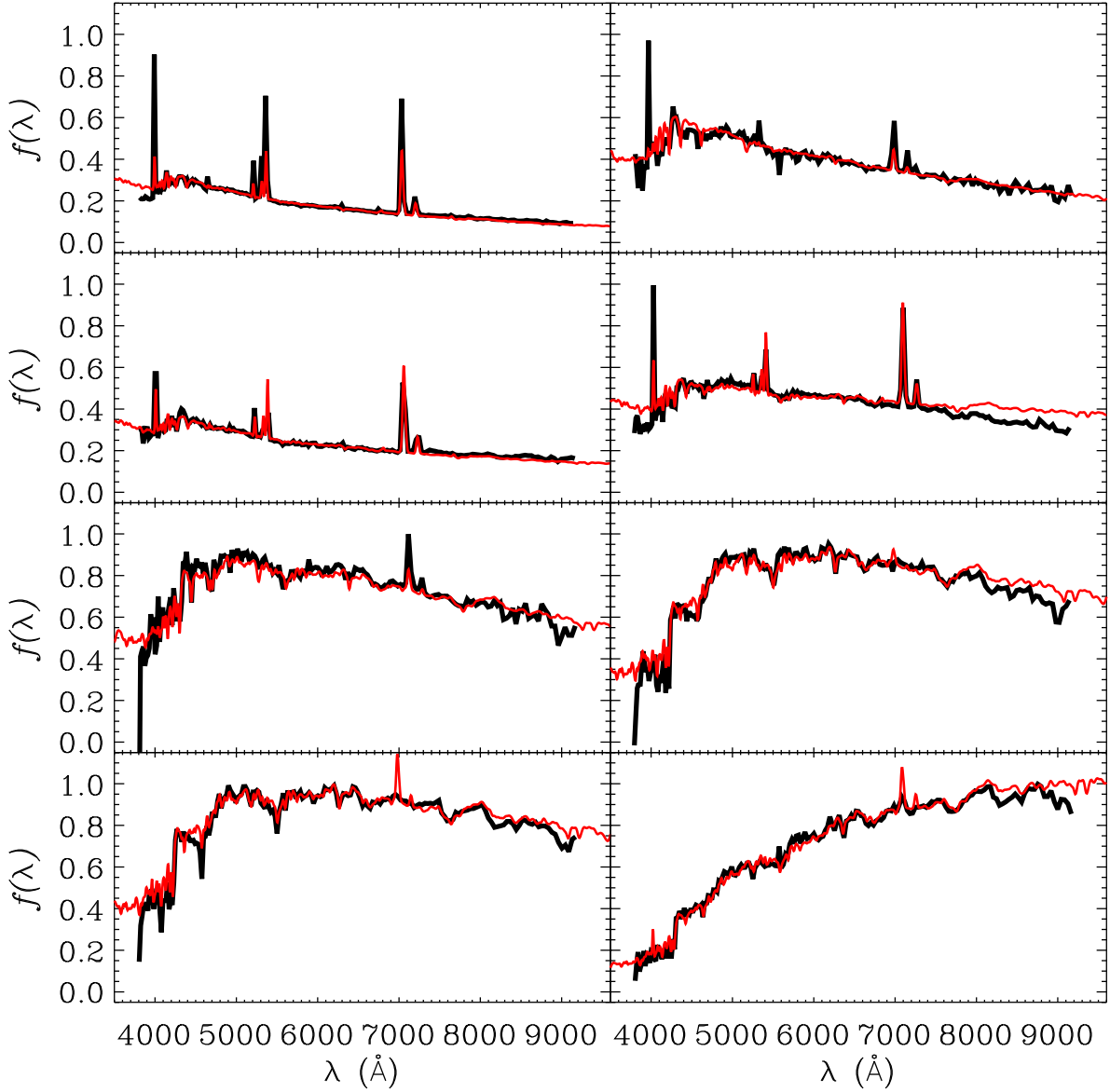}
\caption{\label{specfit} Best fit model spectra based on the five
template fit to $g$, $r$ and $i$ fluxes, compared to the original
SDSS spectra from which we computed those fluxes.  The models and the
original spectra agree very well.}
\end{figure}

We can also test how well these templates recover actual spectra given
only photometry. To do so, we take eight random SDSS spectra (chosen
to span color space), project them onto the $g$, $r$ and $i$
bandpasses, and then fit the five templates to just these three
fluxes.  Figure \ref{specfit} compares the reconstructed model spectra
to the original spectrum. The two agree well in general, though some
features (like the emission lines) are not reproduced well.
Naturally, this success is mostly due to the quality of the original
stellar spectra used in the population synthesis code of
\citet{bruzual03a}. We are not suggesting that $gri$ photometry
contains as much information as the full spectra. However, these
results encourage us that by using broad-band data we can infer
$K$-corrections of galaxies.

\subsection{Predictive power of templates}

Perhaps more interestingly, the templates do a good job of predicting
{\it missing} data. That is, we can ask the question: if we use the
templates to fit only to some bands but leave out others, how well do
the best fits {\it predict} the bands left out?
    
\begin{figure}
\figurenum{\fignum}
\plotone{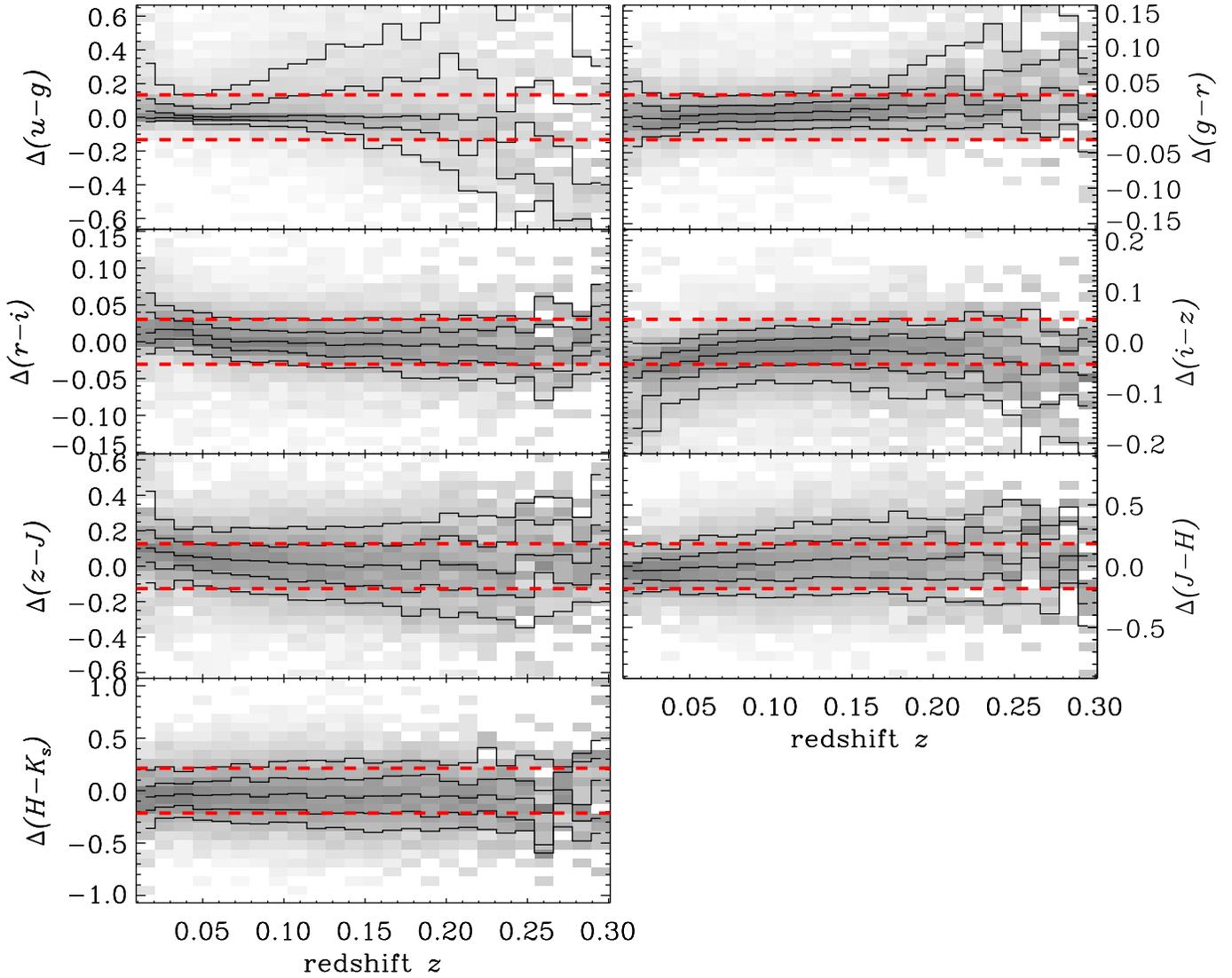}
\caption{\label{twomass_resid} Similar to Figure \ref{fullfits} but
for galaxies observed in both SDSS and 2MASS and only using SDSS and
2MASS bands. The fits are to the SDSS and 2MASS data together.}
\end{figure}

\begin{figure}
\figurenum{\fignum}
\plotone{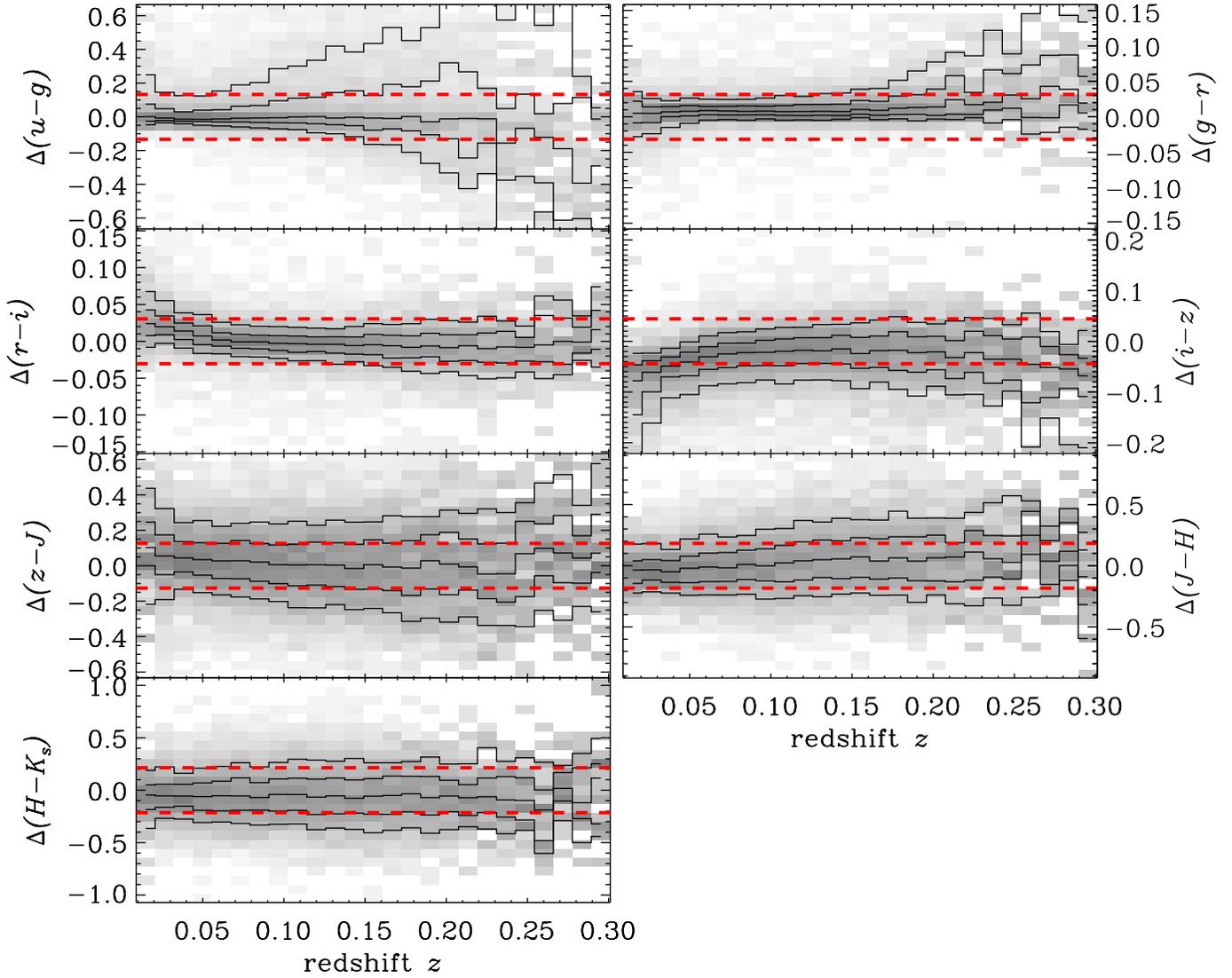}
\caption{\label{twomass_predicted} Similar to Figure
\ref{twomass_resid} but now the fits are {\it only} to the SDSS
bands. The residuals in the 2MASS bands remain very small, indicating
that the 2MASS measurements do not add a lot of information about
these galaxies.}
\end{figure}

Consider Figures \ref{twomass_resid} and \ref{twomass_predicted}. The
former shows the color residuals of {\it fitting} to both SDSS and
2MASS data for each galaxy. The latter shows the color residuals when
we fit {\it only} to the SDSS data and do not include 2MASS in the fit
at all. Without any input from 2MASS the templates do a very good job
of predicting the 2MASS fluxes, with a scatter of 20--30\% --- not far
from the uncertainties in the 2MASS fluxes themselves. Of course, it
is not surprising that it is easy to predict the 2MASS fluxes ---
stellar spectra are very simple in the NIR, so having the $ugriz$-band
fluxes from SDSS yields much information regarding the redder
bandpasses.

However, we will note here that this result flies in the face of a
persistent insistence that NIR observations are necessary to measure
stellar masses of galaxies. If we can predict the NIR observations
themselves, clearly they cannot be adding much to our knowledge of the
underlying stellar mass. This fact changes one's decisions about what
data is best to use for calculating the stellar mass function.  Basing
it on 2MASS data only improves slightly the stellar mass estimate of
each galaxy, while restricting (and thus biasing) the sample
significantly at the lowest luminosities and surface brightnesses
relative to the SDSS. Of course, if at high redshift galaxies have
a very different locus of their SEDs, then perhaps this statement is
less true for them.

\begin{figure}
\figurenum{\fignum}
\plotone{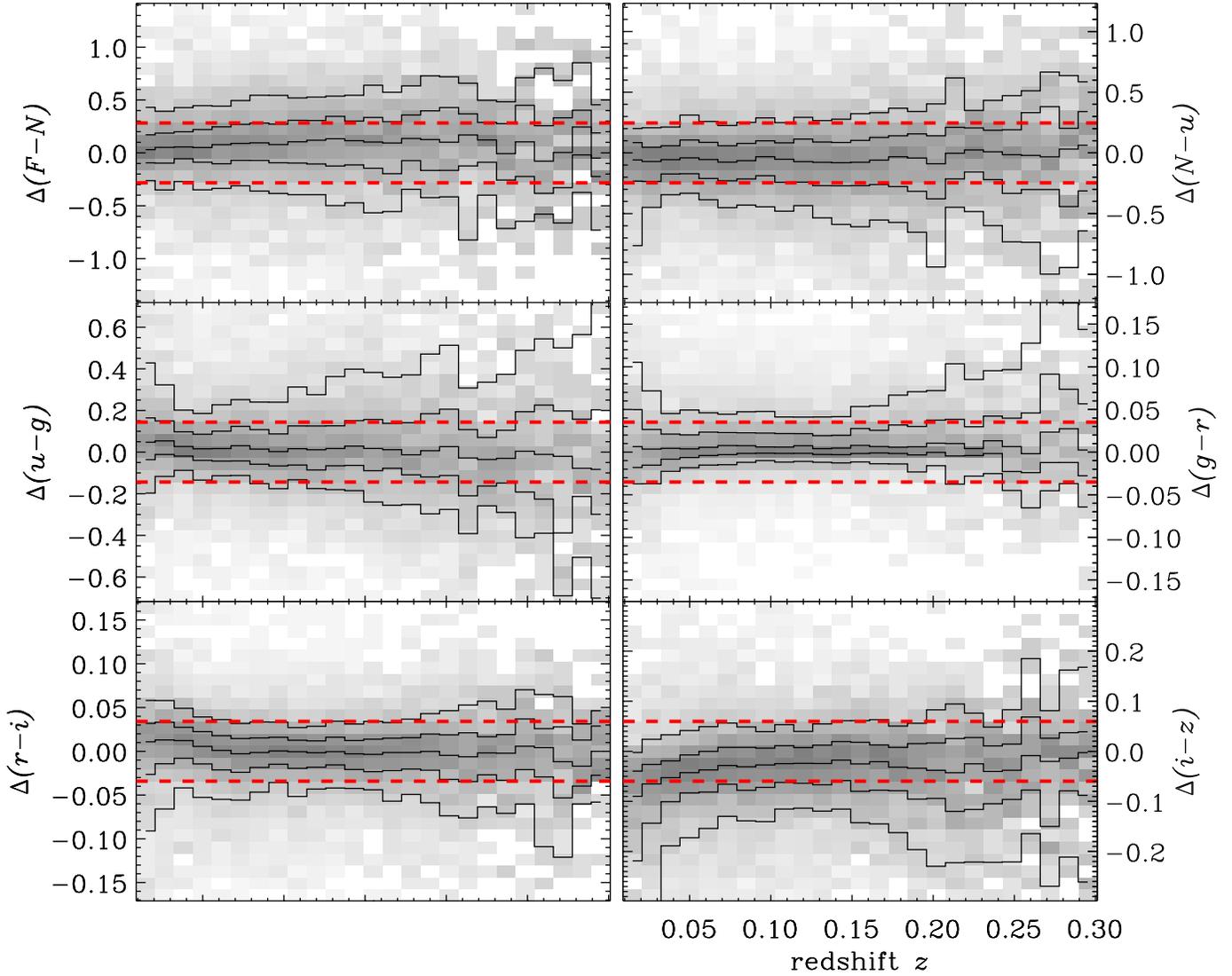}
\caption{\label{galex_resid} Similar to Figure \ref{twomass_resid} but
for galaxies observed in both SDSS and GALEX and only using SDSS and
GALEX bands. The fits are to the SDSS and GALEX data together.}
\end{figure}

\begin{figure}
\figurenum{\fignum}
\plotone{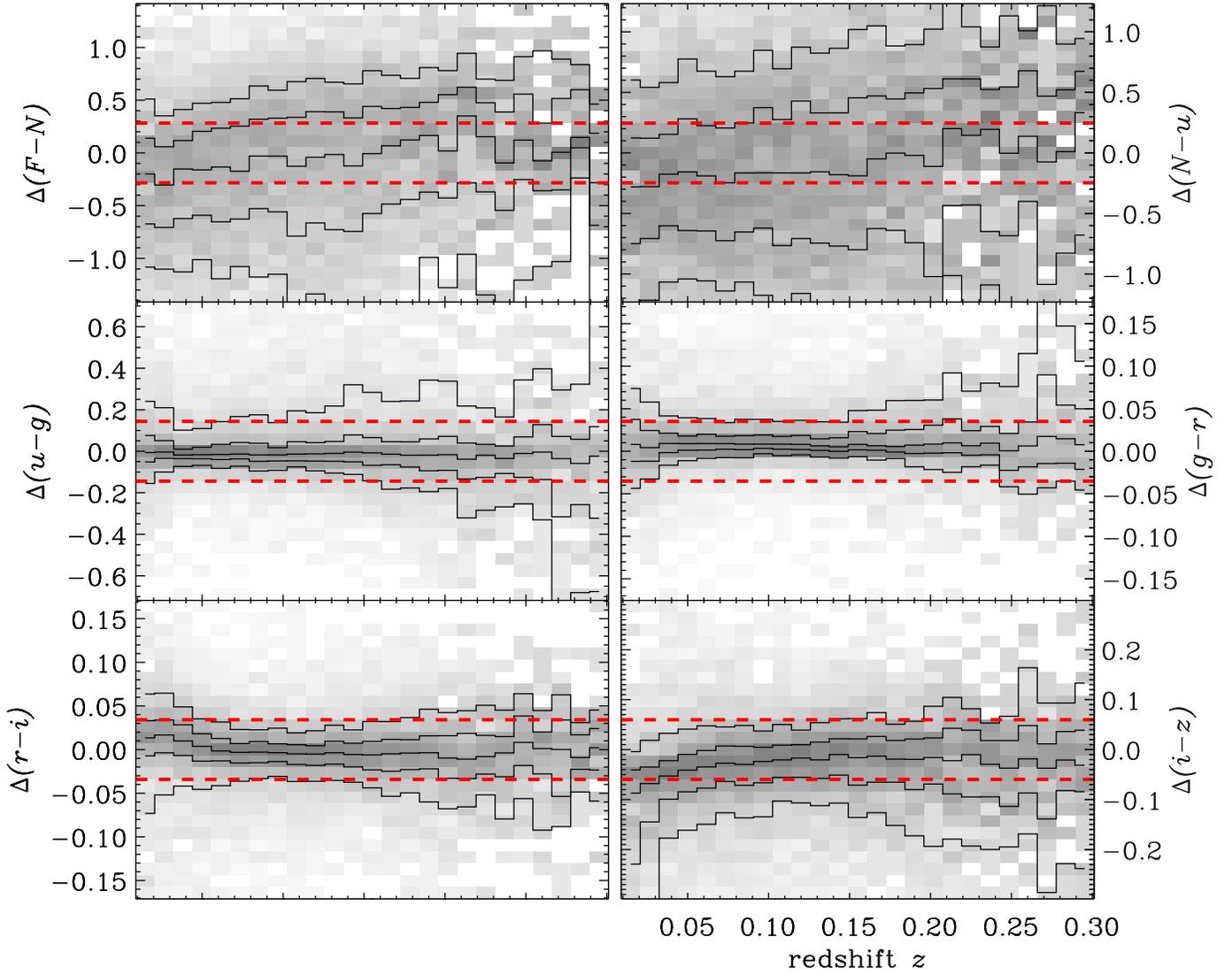}
\caption{\label{galex_predicted} Similar to Figure
\ref{galex_resid} but now the fits are {\it only} to the SDSS
bands. }
\end{figure}

In addition, consider Figures \ref{galex_resid} and
\ref{galex_predicted}. The former shows the color residuals of {\it
fitting} to both SDSS and GALEX data for each galaxy. The latter shows
the color residuals when we fit {\it only} to the SDSS data and do not
include GALEX in the fit at all. Without any input from GALEX the
templates do a significantly worse job at predicting the GALEX
fluxes. The scatter becomes about $\sigma \sim 0.5$ mag.  As is common
knowledge, it is more difficult to predict the UV fluxes, because dust
and recent star-formation are so variable among galaxies. However, the
median residuals based on are our templates still near zero.

\section{Determining $K$-corrections}
\label{kcorrect}

Given a model spectrum for the galaxy, the determination of the
$K$-correction is straightforward. Here we give the relevant formulae
for the $K$-corrections, leaving the derivation to \citet{hogg02a}.
Then, we show the typical $K$-corrections for the data we have fit
to. 

The $K$-correction between a bandpass $R$ used to observe a galaxy at
redshift $z$ and the desired bandpass $Q$ is defined by the equation
(\citealt{oke68a, hogg02a}):
\begin{equation}
\label{kcorrecteqn}
m_R = M_Q + \mathrm{DM}(z) + K_{QR}(z) - 5 \log_{10} h 
\end{equation}
where $\mathrm{DM}(z) = 25 - 5\log_{10} (d_L /
(h^{-1}{\mathrm{~Mpc}}))$ is the bolometric distance modulus
calculated from the luminosity distance $d_L$, and $M_Q$ is the
absolute magnitude. The absolute magnitude is defined as the apparent
magnitude an object would have if were observed 10 pc away, in
bandpass $Q$, at rest.  The traditional definition of the
$K$-correction takes $Q=R$. However, we note that in practice many
surveys do perform $K$-corrections from one observed bandpass $R$ to
another bandpass $Q$ in the rest frame. This practice is particularly
common when dealing with high redshift observations.  In addition to
$K$-corrections, this method also provides an interpretation of the
data in terms of a physical model which describes the stellar mass and
star-formation history of each galaxy.

Equation (\ref{kcorrecteqn}) holds if the $K$-correction $K_{QR}$ is
\begin{equation}
\label{eq:wavelengthL}
K_{QR} = -2.5\,\log_{10}\left[\frac{1}{[1+z]}\,
  \frac{\displaystyle
  \int\mathrm{d}\lambdaobs\,\lambdaobs\,L_{\lambda}\!\left(\frac{\lambdaobs}{1+z}\right)\,R(\lambdaobs)\,
    \int\mathrm{d}\lambdaemit\,\lambdaemit\,
    g^Q_{\lambda}(\lambdaemit)\,Q(\lambdaemit)}
       {\displaystyle
  \int\mathrm{d}\lambdaobs\,\lambdaobs\,g^R_{\lambda}(\lambdaobs)\,R(\lambdaobs)\,
    \int\mathrm{d}\lambdaemit\,\lambdaemit\,
    L_{\lambda}(\lambdaemit)\,Q(\lambdaemit)}
\right] \;\;\;.
\end{equation}
Here, $R(\lambda)$ and $Q(\lambda)$ represent the response of the
instrument per unit photon entering the Earth's atmosphere (or the
telescope aperture for a space instrument).  $g^R_\lambda$ is the flux
density per unit wavelength (e.g.~ ergs s$^{-1}$ cm$^{-2}$ \AA$^{-1}$)
for the standard source for band $R$, and similarly for band $Q$. For
example, if the magnitudes are AB relative, then these represent the
AB standard source, while if they are AB relative then they represent
the spectrum of Vega. 

A particularly common special case is when $R=Q$:
\begin{equation}
\label{eq:specialL}
K_R(z) = -2.5\,\log_{10}\left[\frac{1}{[1+z]}\,
  \frac{\displaystyle
  \int\mathrm{d}\lambdaobs\,\lambdaobs\,L_{\lambda}\!\left(\frac{\lambdaobs}{1+z}\right)\,R(\lambdaobs)\,}
       {\displaystyle
    \int\mathrm{d}\lambdaemit\,\lambdaemit\,
    L_{\lambda}(\lambdaemit)\,R(\lambdaemit)}
\right] \;\;\;.
\end{equation}

\begin{figure}
\figurenum{\fignum}
\plotone{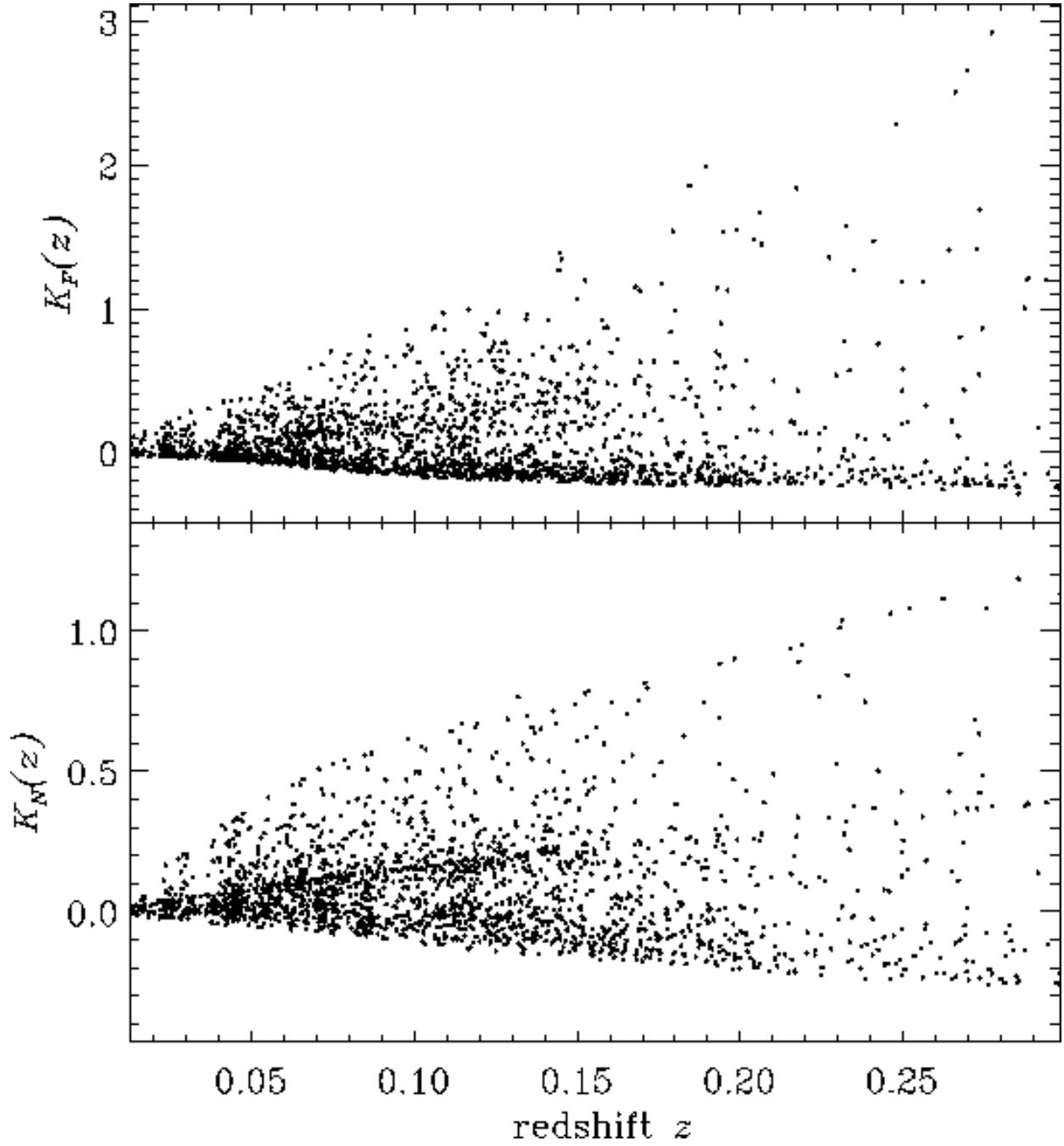}
\caption{\label{galex_kcorrect} $K$-corrections as a function of
redshift in the GALEX near (N) and far (F) UV bands. }
\end{figure}

\begin{figure}
\figurenum{\fignum}
\plotone{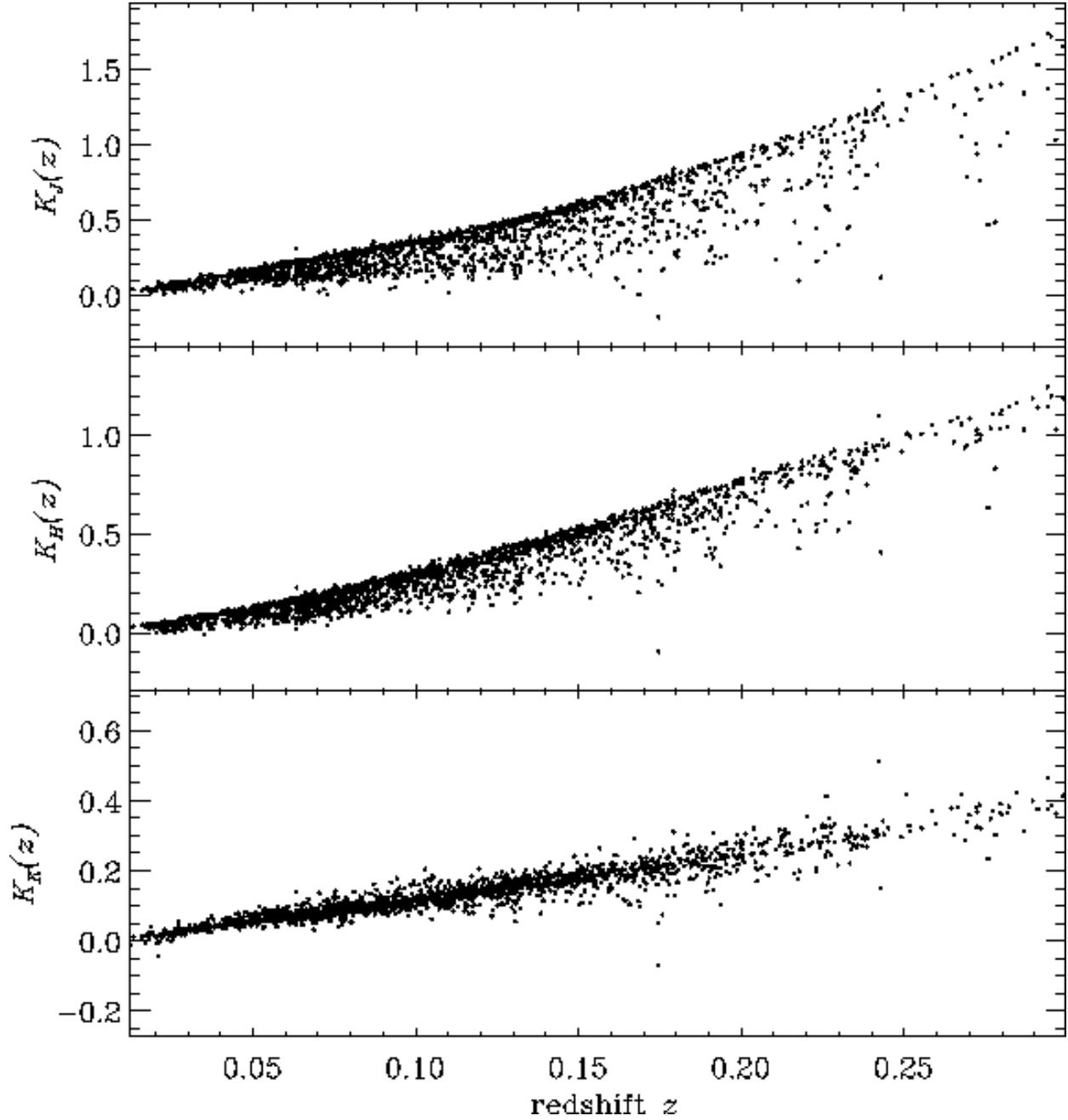}
\caption{\label{twomass_kcorrect} $K$-corrections as a function of
redshift in the 2MASS $J$, $H$ and $K_s$ bands.}
\end{figure}

\begin{figure}
\figurenum{\fignum}
\plotone{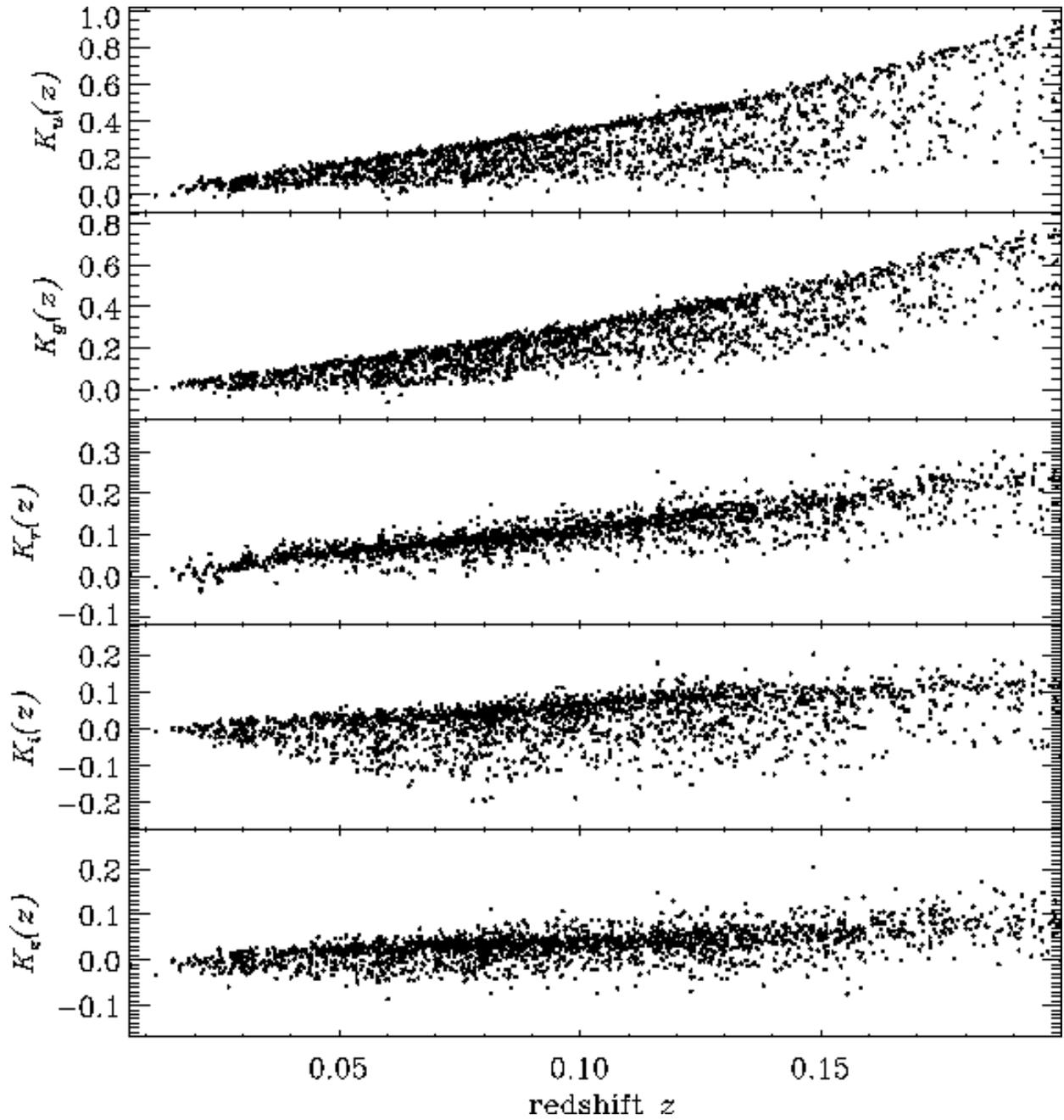}
\caption{\label{sdss_kcorrect} $K$-corrections as a function of
redshift in the SDSS $u$, $g$, $r$, $i$, and $z$ bands.}
\end{figure}

For example, consider Figures \ref{galex_kcorrect},
\ref{sdss_kcorrect} and \ref{twomass_kcorrect}. For a randomly
selected set of galaxies these figures show the $K$-corrections from
the observed frame bandpasses of GALEX, SDSS, and 2MASS respectively,
to the same bandpasses in the rest frame.

Often, it makes sense to $K$-correct to a different set of band passes
than the rest frame. For example, the SDSS Main sample is mostly
around $z=0.1$, so it does not always make sense to $K$-correct from
the observed frame $r$ band to the rest frame $r$ band. Doing so means
applying large and uncertain corrections to all galaxies. If precision
is not important, this uncertainty might be worth the simplification
it brings. However, if one is (say) interested in the evolution of
galaxies, the clustering of galaxies, or something else that requires
precision, it makes more sense to avoid introducing unnecessary
uncertainty.

Our solution to this problem is to correct the magnitude to
blue-shifted bandpasses that correspond to the observed bandpass at
some intermediate redshift. We denote these bandpasses \band{z}{b},
where the blue-shift in this case is by a factor $1+z$. For example,
in the case of SDSS galaxies it makes sense to correct to the
\band{0.1}{r} bandpass --- the $r$ band blue-shifted to $z=0.1$. This
bandpass has the property that the $K$-correction is independent of
the SED of the galaxy. If the magnitudes are AB relative, then the
$K$-correction to \band{z}{b} for a galaxy at redshift $z$ is simply
$-2.5\log_{10} (1+z)$.

\begin{figure}
\figurenum{\fignum}
\plotone{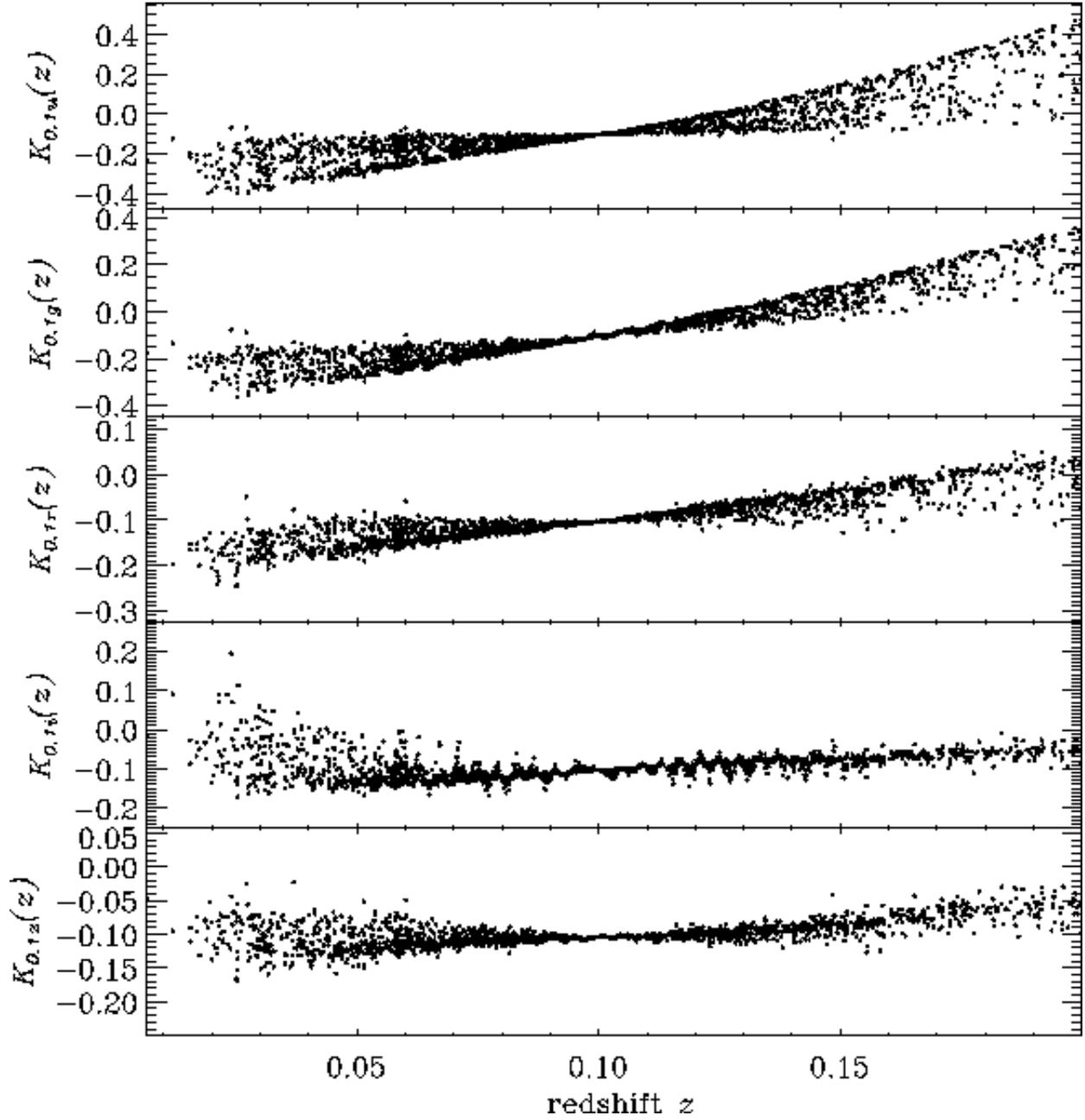}
\caption{\label{main_kcorrect} Same as Figure \ref{sdss_kcorrect}, but
$K$-correcting to the \band{0.1}{u}, \band{0.1}{g}, \band{0.1}{r},
\band{0.1}{i}, and \band{0.1}{z} bands.}  
\end{figure}

Figure \ref{main_kcorrect} shows the $K$-corrections from the
$[ugriz]$ bands to the $\band{0.1}{[ugriz]}$ bands as a function of
redshift for SDSS galaxies. Note that the $K$-corrections converge at
$z=0.1$, so that around that redshift (where most galaxies are) the
uncertainties in the $K$-corrections are minimized. Compare this
simplicity to Figure \ref{sdss_kcorrect}, where the $K$-corrections at
redshift $z=0.1$ in the $g$-band range between $0.$ and $0.2$ (and in
the $u$-band from $0.$ and $0.5$). Most SDSS galaxies are near
redshift $z=0.1$, and obviously the $K$-corrections are not perfect
--- why introduce a large uncertainty for most galaxies when it can
easily be avoided?

\section{Physical interpretation of the models}
\label{physical}

In addition to $K$-corrections, these template fits also provide
physical interpretations of the galaxies, since they correspond to
actual stellar population synthesis models. Indeed, the code outputs
three parameters relevant to the star-formation history: the current
mass of stars in the galaxy; the stellar mass-to-light ratio of the
galaxy; and the fraction of the total amount of star-formation that
has occurred recently. 

\begin{figure}
\figurenum{\fignum}
\plotone{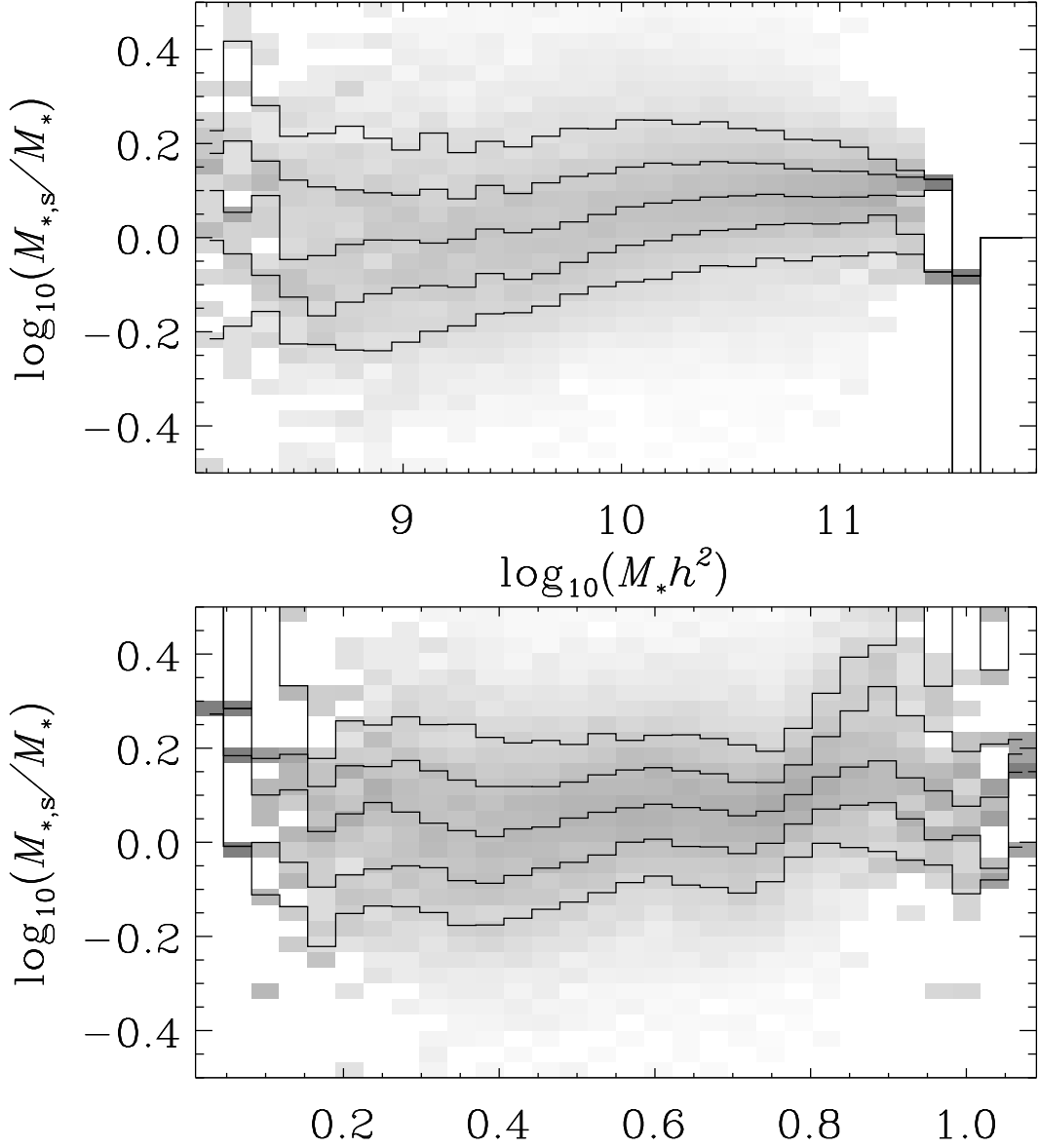}
\caption{\label{mass_to_garching} Our galaxy stellar mass estimates
	$M_\ast$ compared to those of \citet{kauffmann03a} $M_{s,\ast}$, as
	a function of stellar mass (top panel) and of color (bottom
	panel). The greyscale is the conditional distribution
	$M_{s,\ast}/M_\ast$ on each quantity . The lines are the 10\%, 25\%,
	50\%, 75\% and 90\% quantiles.}
\end{figure}

\citet{kauffmann03a} have also calculated stellar masses for SDSS
galaxies. Roughly, they have calculated the $z$-band stellar
mass-to-light ratios of models which best fit the H$\delta$ absorption
and D4000 measurements of the spectra, and then applied those
mass-to-light ratios to the observed $z$-band luminosities of the
imaging. Figure \ref{mass_to_garching} shows a galaxy-by-galaxy
comparison of their stellar masses $M_{\ast,s}$ relative to our
$M_\ast$, as a function of $M_\ast$ and of intrinsic $g-r$ color. The
two sets of masses are very similar to each other, with a scatter of
only 0.1 dex and trends of less than 0.2 dex with stellar mass.  To be
explicit, the masses in Figure \ref{mass_to_garching} correspond to
the current mass in stars remaining in the galaxy. This mass can
differ from the total star-formation rate integrated over time, due to
the fact that stars can die --- and in practice for our fits the total
integrated star-formation rate is usually about twice as high as the
total current mass in stars.

\begin{figure}
\figurenum{\fignum}
\plotone{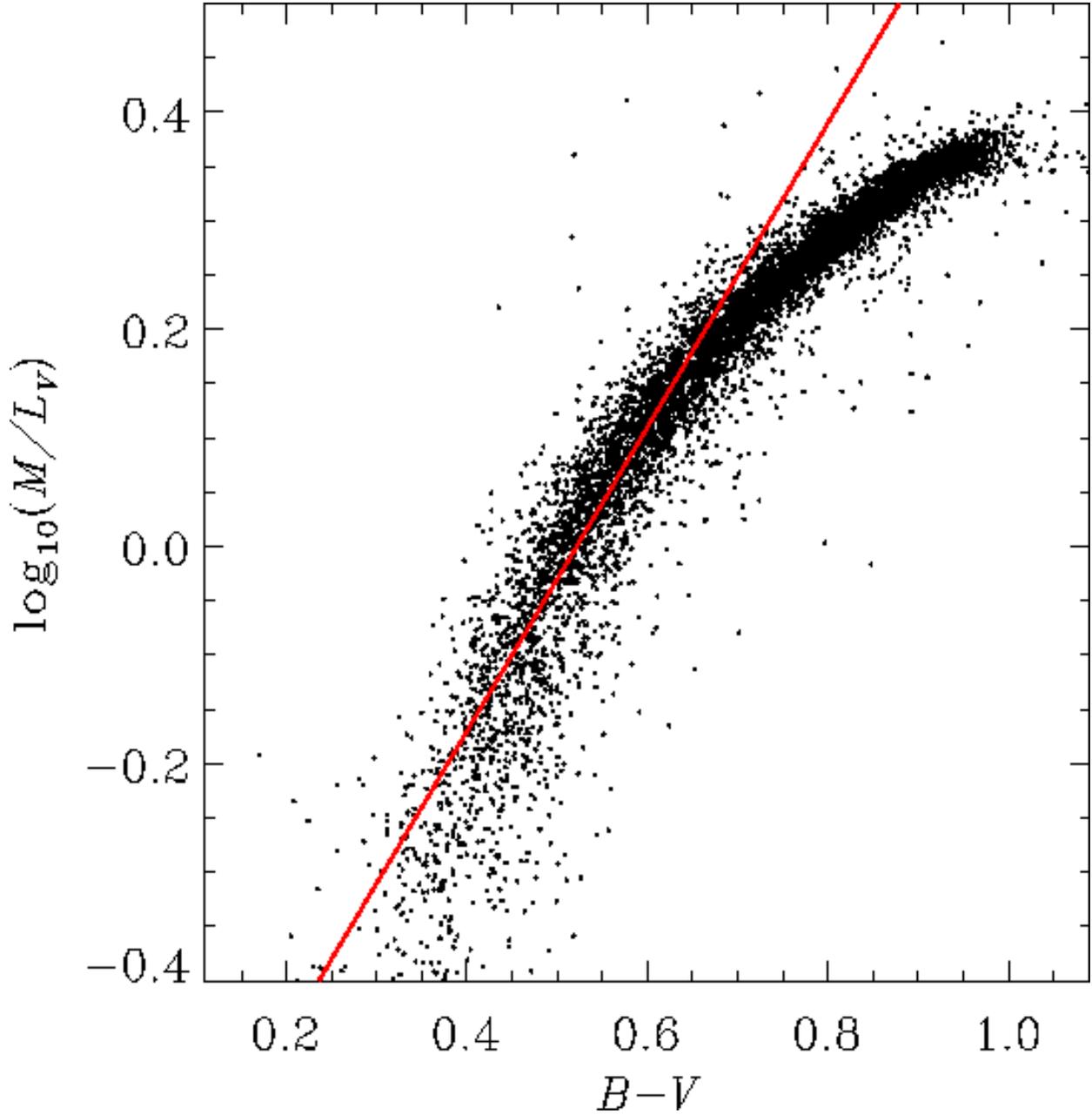}
\caption{\label{mtol} Mass-to-light ratios of galaxies in the $V$ band
	(in solar units) as a function of galaxy $B-V$ color. The solid line
	satisfies the relationship $\log_{10}(M/L_V) = 1.40 (g-r) - 0.73$,
	given by \citet{bell01b} for their sample of spiral galaxies. 
	Their estimates and ours agree for $B-V < 0.8$, where spiral
	galaxies dominate the galaxy population.}
\end{figure}

Similarly, Figure \ref{mtol} shows the mass-to-light ratios of SDSS
galaxies in the $V$ band as a function of color in $B-V$. Here, we
have calculated the $B$ and $V$ band fluxes based on the template fits
to each galaxy, as explained in Section \ref{sdss2bessell} below. We
give the mass-to-light ratio in solar units. The solid line is the
relationship that \citet{bell01b} give based on their fits to spiral
galaxies. For blue galaxies ($B-V < 0.8$) this line is indeed a good
fit to our results. For the typical red galaxy, the \citet{bell01b}
relation predicts a larger stellar mass than we do. 

\begin{figure}
\figurenum{\fignum}
\plotone{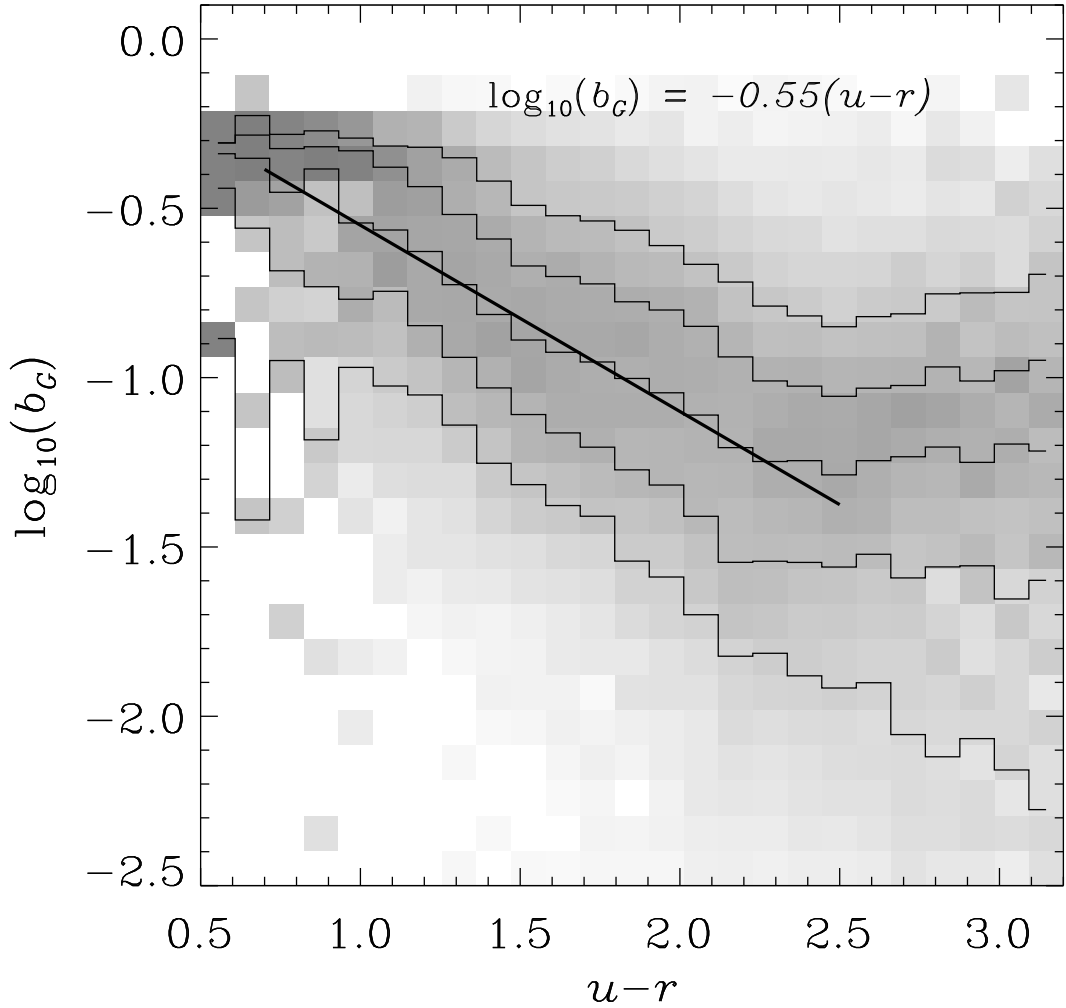}
\caption{\label{umr_bg} Fraction of the total star-formation that has
	occurred in the previous billion years $b_G$, as a function of restframe
	$u-r$ color, for SDSS galaxies. The greyscale is the conditional
	distribution of $b_G$ on $u-r$. The lines are the 10\%, 25\%, 50\%,
	75\% and 90\% quantiles. For $u-r < 2.5$, the median relationship
	follows the simple form listed in the figure. Galaxies with $u-r >
	2.5$ are often highly reddened star-forming galaxies.}
\end{figure}

Finally, we can also calculate measures of the recent star-formation
rate from these fits. Since our fits are to broad-band photometry, and
not based on emission line measurements, we cannot expect to have
detailed information about the very recent star-formation
rate. However, we can try to measure what fraction of the total
star-formation has occurred in, say, the past 1 Gyr:
\begin{equation}
b_{G} = \frac{\int_0^{1\mathrm{~Gyr}} dt\, \mathrm{SFR}(t)}
{\int_0^{a_{\mathrm{Univ}}} dt\, \mathrm{SFR}(t)}
\end{equation}
where the definition here is meant to be similar to the birthrate
parameter $b$ in \citet{kennicutt94a} and references therein.  Note
that here $b_G$ is defined in terms of the total integrated
star-formation history (unlike the stellar masses referred to above)
as is standard for the $b$ parameter. Figure \ref{umr_bg} shows the
relationship between the rest-frame $u-r$ color for SDSS galaxies and
$b_G$. For galaxies with $u-r < 2.5$ (redder than which galaxies are
mostly reddened, edge-on disks) these two parameters follow a simple
relationship:
\begin{equation}
\log_{10} b_g = -0.55 (u-r). 
\end{equation}
Note, however, the large scatter (about 0.3 dex at 1$\sigma$). 

We do not want to overstate the validity of these physical
interpretations. They are sufficiently good physical interpretations
to explain broad band data, but are not uniquely so. Certainly more
detailed spectroscopic analysis can provide better constraints on
star-formation histories and stellar masses. 

One particular weakness for high redshift data is that the templates
always assume that there is 14 Gyrs of star-formation history.  Thus,
{\tt kcorrect} may greatly overestimate the stellar masses of galaxies
at $z \sim 1$ (for example), where the actual star-formation history
must last only 6 Gyrs.

\section{Linear relationships between common magnitude systems}
\label{sdss2bessell}

The software and templates that we present here are also useful for
determining simple conversions between bandpasses performing other
common tasks, such as calculating the absolute magnitude of the Sun in
various band systems, and calculating the conversion between Vega and
AB magnitudes. In this section, we present these tools.

For example, Table \ref{solarmagnitudes} lists the effective
wavelengths, conversion from Vega to AB magnitudes, and absolute
magnitudes of the Sun in a number of filters: the GALEX filters, the
Bessell filters, the SDSS filters, and the 2MASS filters. The listed
numbers are the results of running the IDL functions {\tt
k\_lambda\_eff}, {\tt k\_vega2ab}, and {\tt k\_solar\_magnitudes},
respectively, so the reader can calculate the same thing easily on any
given filter.  The effective wavelengths listed for each filter use
the definition
\begin{equation}
\label{lden}
\lambda_{\mathrm{eff}} = \exp\left[ 
\frac{\int d(\ln\lambda) R(\lambda) \ln\lambda}
{\int d(\ln\lambda) R(\lambda)} \right],
\end{equation}
following \citet{fukugita96a} and \citet{schneider83a}.  For the
spectrum of Vega in the conversion of AB to Vega magnitudes, we use
the \citet{kurucz91a} theoretical Vega spectrum (normalized at 5000
\AA\ to match the \citet{hayes85a} spectrophotometry of Vega). In {\tt
k\_vega2ab}, the user has the option to use the spectrum of
\citet{hayes85a} instead.  We use the solar spectrum of
\citet{kurucz91a} for the calculation of the solar absolute magnitude.

For comparison among results from different surveys, one also wants to
be able to convert the magnitudes in one set of bandpasses to
magnitudes in another set of bandpasses. An example of doing so is
given in our piece of code {\tt sdss2bessell}, which takes SDSS input
magnitudes and outputs absolute magnitudes in the Bessell $U$, $B$,
$V$, $R$ and $I$ bandpasses. 

However, often one wants to just make quick and dirty comparisons,
without going through the trouble of running any software. For these
purposes, we provide Table \ref{lineareqs}, which provides the linear
relationships among the same bandpasses as those listed in Table
\ref{solarmagnitudes}. In fact, these relationships are usually good
to 0.05 mag or better. Note that Table \ref{lineareqs} refers to AB
magnitudes {\it throughout}; use Table \ref{solarmagnitudes} to get
the relationships to Vega magnitudes.

\section{Summary}
\label{summary}

We have here described a method for fitting templates to
nearly-arbitrary sets of spectra and broad-band fluxes. The basic
concept is that we can recover a small set of templates (based on
models), nonnegative linear combinations of which can explain a much
larger set of inhomogeneous observations.  We have applied this method
to SDSS, GALEX, 2MASS, DEEP2, and GOODS data to come up with a set of
templates which are effective for describing all of these data sets.

Nonnegative matrix factorization (NMF) is an effective way to reduce
the dimensionality of any large data set, and has that in common with
PCA.  It is, in some sense, a ``nonnegative PCA.''  However, at least
in the current context the method has several advantages over
PCA. First, it has a natural physical interpretation associated with
that spectral subspace, which is the corresponding subspace of all
possible star-formation histories. Second, it naturally handles data
uncertainties and missing data, which allows it to ignore that
variation which is due purely to statistical errors.  Third, it
handles the complications of observing galaxy spectra photometrically
using broad-band filters. 

We note here that the general NMF method does not {\it depend} on
using a model --- our ``model'' could just have been a set of top hat
functions or Gaussians on a wavelength grid. In some situations, such
as datasets for which there are not well-developed theoretical models,
this approach could be more appropriate.

We have released our results in the form of a templates and a code
base called {\tt kcorrect} which fits those templates to many types of
data (SDSS photometric and spectroscopic data, GALEX data, GOODS data,
DEEP2 data, and 2MASS data). Furthermore, this code returns
$K$-corrections and a physical interpretation of the photometry. All
of the plots in this paper were created using code in the
repository. The code is available from a web page maintained by one of
the authors\footnote{{\tt
http://cosmo.nyu.edu/blanton/kcorrect}}. This web page is kept updated
on improvements in the code and new developments. It consists of a
{\tt C} language library, stand-alone {\tt C} programs, and IDL
language wrappers around the {\tt C} library. Thus, one can use the
basic templates and fitting code using only a machine with a {\tt C}
compiler. However, there is significant functionality which is
programmed in the (unfortunately, proprietary) IDL language. These
routines depend on the {\tt idlutils} library\footnote{{\tt
http://skymaps.info}}.

\acknowledgments

The authors would like to thank Alison Coil, Michael Cooper, David
W. Hogg, John Moustakas, Leonidas Moustakas, David Schiminovich, Risa
Wechsler, and Andrew West for their detailed comments and
feedback. This work was funded by a GALEX Archival Program (38) and an
HST Archival Research grant (AR-9912).

This publication makes use of data products from the Two Micron All
Sky Survey, which is a joint project of the University of
Massachusetts and the Infrared Processing and Analysis
Center/California Institute of Technology, funded by the National
Aeronautics and Space Administration and the National Science
Foundation.

The Galaxy Evolution Explorer (GALEX) is a NASA Small Explorer. The
mission was developed in cooperation with the Centre National d'Etudes
Spatiales of France and the Korean Ministry of Science and Technology.

DEEP2 is a collaboration between UC Santa Cruz and UC Berkeley.
Funding for the DEEP2 survey has been provided by NSF grant
AST-0071048 and AST-0071198.

Funding for the creation and distribution of the SDSS Archive has been
provided by the Alfred P. Sloan Foundation, the Participating
Institutions, the National Aeronautics and Space Administration, the
National Science Foundation, the U.S. Department of Energy, the
Japanese Monbukagakusho, and the Max Planck Society. The SDSS Web site
is {\tt http://www.sdss.org/}.

The SDSS is managed by the Astrophysical Research Consortium (ARC) for
the Participating Institutions. The Participating Institutions are The
University of Chicago, Fermilab, the Institute for Advanced Study, the
Japan Participation Group, The Johns Hopkins University, Los Alamos
National Laboratory, the Max-Planck-Institute for Astronomy (MPIA),
the Max-Planck-Institute for Astrophysics (MPA), New Mexico State
University, Princeton University, the United States Naval Observatory,
and the University of Washington.

\appendix

\section{Nonnegative matrix factorization (NMF)}
\label{nmf}

\subsection{Standard NMF}
The standard non-negative matrix factorization (NMF) problem, as
originally posed by \cite{lee00a}, is to approximate a data
matrix $\XX$ (of size $N_k\times N_n$) as the outer product of two
rank-$N_i$ non-negative matrices $\WW$ and $\HH$. For example, in this
paper, $\XX$ represents the fluxes at $N_n$ different wavelengths for
$N_k$ different galaxies, $\HH$ represents a set of $N_i$ templates to build
each galaxy from, and $\WW$ represents, for each galaxy, the weights
to give each template. This factorization is similar in spirit to
singular value decomposition (SVD) and principal components analysis
(PCA) but crucially, in NMF, $\XX$, $\WW$ and $\HH$ are all
constrained to have non-negative entries. This changes the problem
fundamentally, since no ``cancellation'' of positive and negative
basis functions or coefficients is possible -- all approximation
interactions are strictly additive. The goal, as with SVD or PCA, is
to minmize the squared approximation error (Frobenius norm):
\beqa \label{eq:nmfcost}
\chi^2 &= \|\XX-\WW\HH\|^2\\
&= \sum_{kn} \left(\Xkn - \sum_i \Wki\Hin \right)^2
\eeqa
In their seminal papers, \cite{lee00a} show that this approximation
error in non-increasing under the following very simple multiplicative
update rules: 
\beqa \label{eq:wupdate}
\Wki &\leftarrow \Wki \frac{[\XX\HH\T]_{ki}}{[\WW\HH\HH\T]_{ki}}\\
\label{eq:hupdate}
\Hin &\leftarrow \Hin \frac{[\WW\T\XX]_{in}}{[\WW\T\WW\HH]_{in}}
\eeqa
$[M]_{ab}$ denotes the $(ab)$ element of the matrix valued expression
$M$. $\HH\T$ is the transpose of matrix $\HH$.

These rules are remarkable because, although they make finite (not
infinitesimal) adjustments to the elements of the approximation
matrices, they have no step-size parameters and are always guaranteed
to reduce the error (or leave it invariant once they have converged).
$\chi^2$ is in general a ``non-convex'' function of $\WW$ and $\HH$,
meaning we cannot guarantee there is only one local
minimum. Therefore, the procedure does not necessarily find the global
optimum. But in practice, even with random initialization, these rules
seem to converge to good solutions for real data (at least data like
ours).

\subsection{NMF in the space of coefficients for a known basis}
The factorization problem we face in this paper is a slight variation on the
basic NMF setup described above. We wish to approximate $\XX$ by the
product $\AA\BB\MM$, 
where $\MM$ is a given (fixed) basis matrix of size $N_j \times N_n$
and the optimization is over the matrices $\AA$ (of size
$N_k\times N_i$) and $\BB$ (of size $N_i\times N_j$). All matrices
$\XX,\AA,\BB\,\MM$ have non-negative entries. This can be thought of
as performing non-negative matrix factorization on the coefficients of
an approximation, given a fixed (non-negative) basis, described by the
columns of $\MM$. Once again, the objective we wish to minimize is the
squared approximation error: 
\beqa 
\chi^2 &= \|\XX-\AA\BB\MM\|^2\\
&= \sum_{kn} \left(\Xkn - \sum_{ij} \Aki\Bij\Mjn \right)^2
\eeqa
This equation is almost that posed in Equation \ref{chi2}.
In the case of this paper, this generalization allows us to express
the templates in terms of star-formation histories, but to compare the
predicted fluxes for the star-formation histories through the $\MM$
matrix to the observed fluxes $\XX$.

Of course, in the very special case where $\MM$ is an invertible
matrix, this problem can be transformed into the original NMF problem above by
right multiplying the data matrix by $\MM\inv$. However, in most
situations, including ours, $\MM$ has many fewer rows than columns, and
as such is far from invertible. Fortunately, however, it is possible
to derive multiplicative updates for this extended problem 
which minimze the error directly, even when $\MM$ is not invertible.

First, by thinking of the tensor product $\BB\MM$ as a single
non-negative matrix $\HH$, we can trivially derive a multiplicative
update equation for the elements of $\AA$ by using the $\WW$ update
provided above: 
\beq
\Aki \leftarrow \Aki 
\frac{[\XX\MM\T\BB\T]_{ki}}{[\AA\BB\MM\MM\T\BB\T]_{ki}}
\eeq
By Lee and Seung's original proof, this update is guaranteed (for any
non-negative matrices $\XX$,$\MM$ and $\BB$) not to increase the
approximation error. 

Our main algorithmic contribution is to derive a similar update equation
for the elements of $\BB$:
\beq \label{eq:bupdate}
\Bij \leftarrow \Bij
\frac{[\AA\T\XX\MM\T]_{ij}}{[\AA\T\AA\BB\MM\MM\T]_{ij}}
\eeq
Following the technique outlined in \cite{lee00a}, it can still be shown
that the error is nonincreasing under the application of this update.

The proof involves the use of an inequality lemma for symmetric
non-negative matrices:

{\bf Lemma:} 
{\it For any symmetric matrix $\PP$ having non-negative entries $\Pni
\geq 0$, any vector $\zz$ having non-negative entries $\zn \geq 0$,
and any vector $\yy$,}
\beq
\sum_{ni} \yn \yi \Pni \leq \sum_{ni} \yn^2 \frac{\zi}{\zn} \Pni 
\eeq

\begin{figure}
\figurenum{\fignum}
\plotone{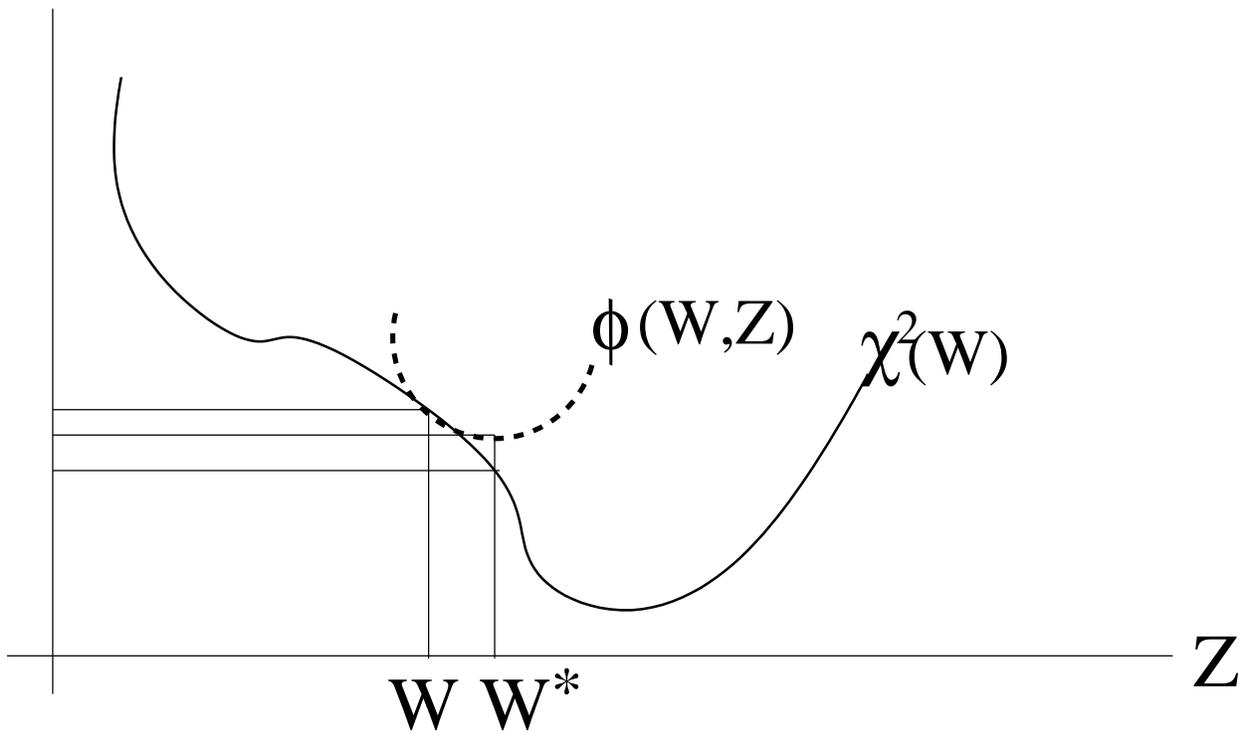}
\caption{\label{lemma} Example diagram of definition of $\phi$ in the
	text. $\chi^2$ represents the actual $\chi^2$ while $\phi$ is
	designed such that it exceeds $\chi^2$ except at the single point
	$W$. By minimizing $\phi$ we can find a point $W^\ast$ that we know
	to have an equal or better $\chi^2$ than that at $W$. }
\end{figure}

For the standard NMF problem, we can prove that the approximation
error \eqref{eq:nmfcost} is non-increasing under \eqref{eq:wupdate} by
using the above lemma to construct a function $\phi(\WW,\ZZ)$ which is
an upper bound on the cost $\chi^2(\WW)$ for any non-negative matrix $\ZZ$:
{\small 
\beqa 
\chi^2(\WW) &= \sum_{kni\ell} \Wki\Wkl \Hin\Hln
- 2 \sum_{kni} \Wki \Xkn \Hin + \sum_{kn} \Xkn^2 \\
\phi(\WW,\ZZ) &= \sum_{kni\ell} \Wki^2\frac{\Zkl}{\Zki} \Hin\Hln 
- 2 \sum_{kni} \Wki \Xkn \Hin + \sum_{kn} \Xkn^2 \\
\phi(\WW,\ZZ) &\geq \chi^2(\WW) \qquad \forall \: \Wki\geq 0,\: \Zki\geq 0
\eeqa
} with equality being achieved when $\ZZ=\WW$. Figure \ref{lemma}
represents this definition schematically, showing the true $\chi^2$
function and $\phi$. The trick of the method is to define $\phi$ such
that it is equal to $\chi^2$ at $W$, greater than $\chi^2$ everywhere
else, and is easily minimizable. Then one can use the minimum of
$\phi$ as the update, and be guaranteed that one's updates do not
increase $\chi^2$.

The function $\phi$ can be analytically minimized with respect to
its first argument:
\beqa
\phi(\WW^*,\ZZ) &\leq \phi(\WW,\ZZ) \qquad \forall \: \WW,\ZZ \\
\Wki^* &= \Zki \frac{\sum_n \Xkn \Hin}{\sum_{n\ell} \Hin\Hln \Zkl}
\eeqa
If we set $\ZZ=\WW$, this becomes exactly \eqref{eq:wupdate} and
now we can easily prove the validity of this update rule: 
\beq
\chi^2(\WW) = \phi(\WW,\WW) \geq \phi(\WW^*,\WW) \geq \chi^2(\WW^*)
\eeq
where the first inequality comes from the fact that $\WW^*$ minimizes
$\phi(\cdot,\WW)$ with respect to its first argument
and the second comes from the fact that $\phi(\WW^*,\cdot)$ is a bound
on $\chi^2(\WW^*)$. The proof of validity for the update
\eqref{eq:hupdate} is analogous by symmetry. 

To prove the validity of our update \eqref{eq:bupdate} for our new
problem, we proceed in a similar fashion, using the lemma twice to
construct consecutive upper bounds on the cost:
{\small 
\beqa 
\chi^2(\BB) &= \sum_{kni\ell{}js} \Aki\Akl\Bij\Bls\Mjn\Msn\\ \nonumber
&- 2 \sum_{knij} \Xkn\Aki\Bij\Mjn + \sum_{kn} \Xkn^2 \\
\phi(\BB,\ZZ,\beta) &= \sum_{kni\ell{}js} \Aki\Akl\Bij^2
\frac{\Zks\betalj}{\Zkj\betaij}\Mjn\Msn\\ \nonumber
&- 2 \sum_{knij} \Xkn\Aki\Bij\Mjn + \sum_{kn} \Xkn^2 \\
\phi(\BB,\ZZ,\beta) &\geq \chi^2(\BB) \qquad \forall 
\: \Bij\geq 0,\: \Zkj\geq 0,\: \betaij\geq 0
\eeqa
} with equality being achieved when $\ZZ=\AA\BB$ and 
$\beta=\BB$.

Once again, the bound $\phi$ can be analytically minimized with respect to
its first argument:
\beqa
\phi(\BB^*,\ZZ,\beta) &\leq \phi(\BB,\ZZ,\beta) \qquad 
\forall \: \BB,\ZZ,\beta \\
\Bij^* &= \betaij \frac{\sum_{kn} \Xkn\Aki\Mjn}
{\sum_{kn\ell{}s} \Aki\Akl\frac{\Zks\betalj}{\Zkj}\Mjn\Msn}
\eeqa
If we set $\ZZ=\AA\BB$ and $\beta=\BB$, this becomes
exactly \eqref{eq:bupdate} and 
we can now prove the validity of this update rule: 
\beq
\chi^2(\BB) = \phi(\BB,\AA\BB,\BB) \geq \phi(\BB^*,\AA\BB,\BB) 
\geq \chi^2(\BB^*) 
\eeq
where the first inequality comes from the fact that $\BB^*$ minimizes
$\phi(\cdot,\AA\BB,\BB)$ with respect to its first argument
and the second comes from the fact that $\phi(\BB^*,\cdot,\cdot)$ is a bound
on $\chi^2(\BB^*)$. The only condition we require is that the elements
of the matrix $\MM\MM\T$ are all non-negative.

\subsection{Nonuniform uncertainties}
When different entries $\Xkn$ of the data matrix have different
uncertainties, the natural objective function is the weighted approximation
error (which also corresponds to the negative log likelihood under a
Gaussian noise assumption):
\beq
\chi^2 = \sum_{kn} \left( \frac{\Xkn - 
\sum_i \Wki\Hin}{\skn}\right)^2
\eeq

This case can easily be handled since during the update for $\WW$ the
elements of $\HH$ are fixed and vice versa and the updates are
guaranteed not to increase the cost for any nonnegative matrices.
In particular, when updating $\WW$, we can rewrite the cost function as
\beq
\chi^2 = \sum_{kn} \left( \frac{\Xkn}{\skn} - 
\sum_i \Wki\frac{\Hin}{\skn}\right)^2
\eeq
yielding updates of the form:
\beq
\Wki \leftarrow \Wki 
\left( \sum_n \frac{\Xkn\Hin}{\skn^2} \right) / 
\left( \sum_{mn} \frac{W_{km}H_{mn}\Hin}{\skn^2}\right)
\eeq

and similarly, when updating $\HH$, we can rewrite the cost function as
\beq
\chi^2 = \sum_{kn} \left( \frac{\Xkn}{\skn} - 
\sum_i \frac{\Wki}{\skn}\Hin\right)^2
\eeq
yielding updates of the form:
\beq
\Hin \leftarrow \Hin
\left( \sum_k \frac{\Wki\Xkn}{\skn^2}\right) /
\left( \sum_{mk} \frac{\Wki{}W_{km}H_{mn}}{\skn^2}\right)
\eeq

This argument can be equally applied to our extended model, yielding
the final update equations which are actually implemented in the
kcorrect code:
\beqa 
\Aki &\leftarrow \Aki 
\left( \sum_{jn} \frac{\Xkn\Mjn\Bij}{\skn^2}\right) / 
\left( \sum_{mljn} \frac{A_{km}B_{mj}M_{jn}M_{nl}B_{il}}{\skn^2}\right) \\
\Bij &\leftarrow \Bij
\left( \sum_{kn} \frac{\Aki\Xkn\Mjn}{\skn^2}\right) / 
\left( \sum_{mlkn} \frac{A_{ki}A_{km}B_{ml}M_{ln}M_{jn}}{\skn^2}\right)
\eeqa


\section{Format for templates}
\label{format}

We have fit several different sets of templates that we release
with the code, which we denote:
\begin{enumerate}
\item {\tt default}, the default set of five templates, 
\item {\tt lrg1}, the single template fit to LRGs, and 
\item {\tt goods}, the five template fit to just the GOODS data.
\end{enumerate}
The information about the default set of templates is contained in the
file {\tt data/templates/k\_nmf\_derived.default.fits} in the {\tt
kcorrect} project.  This file has 25 Header and Data Units (HDUs),
each listed in Table \ref{derived} and described in more detail in the
paragraphs below.  There are similar files for the {\tt lrg1} and {\tt
goods} template sets.

As described in Section \ref{models}, there are $N_{\mathrm{basis}} =
485$ basis spectra, consisting of 450 different instantaneous burst
stellar populations ($N_{\mathrm{age}} = 25$ ages, $N_{\mathrm{mets}}
= 6$ metallicities, and $N_{\mathrm{dust}}=3$ dust properties) plus 35
different emission line models. The {\tt templates} HDU has the
coefficients of the five templates in this basis space. 

For each of the five templates, we have the template spectrum {\tt
spec} in units of ergs s$^{-1}$ cm$^{-2}$ \AA$^{-1}$ per solar mass,
as it would be observed at 10 pc distance. The wavelength grid is in
the {\tt lambda} HDU. As noted above, the models are smoothed at 300
km s$^{-1}$ resolution; the {\tt rawspec} HDUs have the original
(unsmoothed) models from \citet{bruzual03a}. In addition, we give
versions without the emission lines (with the {\tt \_nl} suffix) and
without the dust extinction applied (with the {\tt \_nd} suffix). The
actual emission-line-only spectra for each template are in the {\tt
lspec} HDU. Finally, the multiplicative dust extinction factor for
each template spectrum is given in the {\tt extinction} HDU.

We include the star-formation rate as a function of time in the {\tt
sfr} HDU (the age grid used is in the {\tt ages} HDU). The time
differential used to quantify this rate is in the {\tt dage} HDU; to
get the total number of stars formed at each age, multiply {\tt sfr}
by {\tt dage}. The average metallicity of the stars formed as a
function of time is in the {\tt metallicity} HDU.

The total mass formed in each template (in solar masses) is in the
{\tt mass} HDU --- this is actually just unity for each template. The
total surviving stellar mass for each template is in the {\tt mremain}
HDU. The metallicity in the surviving stars is given in the {\tt mets}
HDU.  The total mass in stars formed in the last 300 Myrs is in the
{\tt m300} HDU, and the total mass formed in the last 1 Gyr is in the
{\tt m1000} HDU.

The properties of the 450 stellar population basis vectors are given
in the four HDU: {\tt basis\_ages}, {\tt basis\_mets}, {\tt
basis\_dusts}, and {\tt basis\_mremain}. Respectively, these give the
ages, metallicities, dust properties, and fraction of original stellar
mass remaining for each instantaneous burst in the grid. The dust 
properties are given in terms of a structure with four elements which
refer to the properties in the multiple scattering model of \citet{witt00a}:
\begin{enumerate}
\item {\tt GEOMETRY}: the geometry of the dust (e.g. ``shell'') 
\item {\tt DUST}: the dust extinction curve (e.g. ``MW'' for Milky
	Way extinction and ``SMC'' for Small Magellanic Cloud type
	extinction) 
\item {\tt STRUCTURE}: structure of the dust distribution (``h'' for
	homogeneous, ``c'' for clumpy)
\item {\tt TAUV}: the amount of dust (the total optical depth in the
	$V$ band)
\end{enumerate}

\newpage

\clearpage
\begin{deluxetable}{rrrrr}
\tablewidth{0pt}
\tablecolumns{5}
\tablecaption{\label{solarmagnitudes} Properties of various filters}
\tablehead{ Band & $\lambda_{\mathrm{eff}}$ (\AA) & $m_{\mathrm{AB}} -
	m_{\mathrm{Vega}}$ & $M_\odot$ (AB) &
	$M_\odot$ (Vega)}
\startdata
$U$ & 3571. & 0.79 & 6.35 & 5.55\cr
$B$ & 4344. & -0.09 & 5.36 & 5.45\cr
$V$ & 5456. & 0.02 & 4.80 & 4.78\cr
$R$ & 6442. & 0.21 & 4.61 & 4.41\cr
$I$ & 7994. & 0.45 & 4.52 & 4.07\cr
$u$ & 3546. & 0.91 & 6.38 & 5.47\cr
$g$ & 4670. & -0.08 & 5.12 & 5.20\cr
$r$ & 6156. & 0.16 & 4.64 & 4.49\cr
$i$ & 7472. & 0.37 & 4.53 & 4.16\cr
$z$ & 8917. & 0.54 & 4.51 & 3.97\cr
$J$ & 12355. & 0.91 & 4.56 & 3.65\cr
$H$ & 16458. & 1.39 & 4.71 & 3.32\cr
$K_s$ & 21603. & 1.85 & 5.14 & 3.29\cr
\band{0.1}{u} & 3224. & 1.25 & 6.78 & 5.53\cr
\band{0.1}{g} & 4245. & -0.01 & 5.43 & 5.44\cr
\band{0.1}{r} & 5597. & 0.04 & 4.76 & 4.71\cr
\band{0.1}{i} & 6792. & 0.27 & 4.57 & 4.30\cr
\band{0.1}{z} & 8107. & 0.46 & 4.52 & 4.05\cr
\enddata
\tablecomments{Uses model solar spectrum and model Vega spectrum of
	\citet{kurucz91a}. Effective wavelength is defined in the text. The
	$UBRVI$ filters are those of \citet{bessell90a}. The $ugriz$ filters
	are those determined by Mamoru Doi, Daniel Eisenstein, and James
	Gunn and available on the SDSS DR4 web site.\footnote{{\tt
	http://www.sdss.org/dr4/}} The $JHK_s$ filters are those from
	\citet{cohen03a}. }
\end{deluxetable}

\begin{deluxetable}{lr}
\tablewidth{0pt}
\tablecolumns{2}
\tablecaption{\label{lineareqs} Conversions among various filters}
\tablehead{ Equation & Color dispersion }
\startdata
$u = \band{0.1}{u} - 0.3310 - 0.3014 \left[ (\band{0.1}{u}-\band{0.1}{g}) - 1.2839 \right] $ & $\sigma\left[\band{0.1}{u}-\band{0.1}{g}\right] = 0.28$ \cr
$g = \band{0.1}{g} - 0.3112 - 0.3530 \left[ (\band{0.1}{g}-\band{0.1}{r}) - 0.7187 \right] $ & $\sigma\left[\band{0.1}{g}-\band{0.1}{r}\right] = 0.20$ \cr
$r = \band{0.1}{r} - 0.2026 - 0.3810 \left[ (\band{0.1}{r}-\band{0.1}{i}) - 0.3530 \right] $ & $\sigma\left[\band{0.1}{r}-\band{0.1}{i}\right] = 0.07$ \cr
$i = \band{0.1}{i} - 0.1072 - 0.4273 \left[ (\band{0.1}{i}-\band{0.1}{z}) - 0.1914 \right] $ & $\sigma\left[\band{0.1}{i}-\band{0.1}{z}\right] = 0.13$ \cr
$z = \band{0.1}{i} - 0.3161 - 1.2057 \left[ (\band{0.1}{i}-\band{0.1}{z}) - 0.1914 \right] $ & $\sigma\left[\band{0.1}{i}-\band{0.1}{z}\right] = 0.13$ \cr
$u = U + 0.0682 + 0.0197 \left[ (U-B) - 0.9602 \right] $ & $\sigma\left[U-B\right] = 0.20$ \cr
$g = B - 0.2354 - 0.3411 \left[ (B-V) - 0.5870 \right] $ & $\sigma\left[B-V\right] = 0.18$ \cr
$r = V - 0.2585 - 0.5003 \left[ (V-R) - 0.3161 \right] $ & $\sigma\left[V-R\right] = 0.07$ \cr
$i = R - 0.2000 - 0.4248 \left[ (R-I) - 0.2652 \right] $ & $\sigma\left[R-I\right] = 0.12$ \cr
$z = R - 0.4088 - 1.2495 \left[ (R-I) - 0.2652 \right] $ & $\sigma\left[R-I\right] = 0.12$ \cr
$\band{0.1}{u} = U + 0.3989 + 0.4135 \left[ (U-B) - 0.9602 \right] $ & $\sigma\left[U-B\right] = 0.20$ \cr
$\band{0.1}{g} = U - 0.8845 - 0.9508 \left[ (U-B) - 0.9602 \right] $ & $\sigma\left[U-B\right] = 0.20$ \cr
$\band{0.1}{g} = B + 0.0759 + 0.0620 \left[ (B-V) - 0.5870 \right] $ & $\sigma\left[B-V\right] = 0.18$ \cr
$\band{0.1}{r} = B - 0.6429 - 1.0845 \left[ (B-V) - 0.5870 \right] $ & $\sigma\left[B-V\right] = 0.18$ \cr
$\band{0.1}{r} = V - 0.0558 - 0.1803 \left[ (V-R) - 0.3161 \right] $ & $\sigma\left[V-R\right] = 0.07$ \cr
$\band{0.1}{i} = R - 0.0927 + 0.0035 \left[ (R-I) - 0.2652 \right] $ & $\sigma\left[R-I\right] = 0.12$ \cr
$\band{0.1}{z} = R - 0.2841 - 1.0301 \left[ (R-I) - 0.2652 \right] $ & $\sigma\left[R-I\right] = 0.12$ \cr
$\band{0.1}{u} = u + 0.3310 + 0.3203 \left[ (u-g) - 1.2638 \right] $ & $\sigma\left[u-g\right] = 0.26$ \cr
$\band{0.1}{g} = u - 0.9528 - 0.7572 \left[ (u-g) - 1.2638 \right] $ & $\sigma\left[u-g\right] = 0.26$ \cr
$\band{0.1}{g} = g + 0.3113 + 0.4620 \left[ (g-r) - 0.6102 \right] $ & $\sigma\left[g-r\right] = 0.15$ \cr
$\band{0.1}{r} = g - 0.4075 - 0.8577 \left[ (g-r) - 0.6102 \right] $ & $\sigma\left[g-r\right] = 0.15$ \cr
$\band{0.1}{i} = r - 0.1504 - 0.3654 \left[ (r-i) - 0.2589 \right] $ & $\sigma\left[r-i\right] = 0.10$ \cr
$\band{0.1}{z} = i - 0.0836 - 0.7518 \left[ (i-z) - 0.2083 \right] $ & $\sigma\left[i-z\right] = 0.10$ \cr
$U = u - 0.0682 - 0.0140 \left[ (u-g) - 1.2638 \right] $ & $\sigma\left[u-g\right] = 0.26$ \cr
$B = u - 1.0286 - 0.7981 \left[ (u-g) - 1.2638 \right] $ & $\sigma\left[u-g\right] = 0.26$ \cr
$B = g + 0.2354 + 0.3915 \left[ (g-r) - 0.6102 \right] $ & $\sigma\left[g-r\right] = 0.15$ \cr
$V = g - 0.3516 - 0.7585 \left[ (g-r) - 0.6102 \right] $ & $\sigma\left[g-r\right] = 0.15$ \cr
$R = r - 0.0576 - 0.3718 \left[ (r-i) - 0.2589 \right] $ & $\sigma\left[r-i\right] = 0.10$ \cr
$I = i - 0.0647 - 0.7177 \left[ (i-z) - 0.2083 \right] $ & $\sigma\left[i-z\right] = 0.10$ \cr
$U = \band{0.1}{u} - 0.3992 - 0.3189 \left[ (\band{0.1}{u}-\band{0.1}{g}) - 1.2839 \right] $ & $\sigma\left[\band{0.1}{u}-\band{0.1}{g}\right] = 0.28$ \cr
$B = \band{0.1}{g} - 0.0759 - 0.0545 \left[ (\band{0.1}{g}-\band{0.1}{r}) - 0.7187 \right] $ & $\sigma\left[\band{0.1}{g}-\band{0.1}{r}\right] = 0.20$ \cr
$V = \band{0.1}{g} - 0.6628 - 0.9259 \left[ (\band{0.1}{g}-\band{0.1}{r}) - 0.7187 \right] $ & $\sigma\left[\band{0.1}{g}-\band{0.1}{r}\right] = 0.20$ \cr
$R = \band{0.1}{r} - 0.2603 - 0.9162 \left[ (\band{0.1}{r}-\band{0.1}{i}) - 0.3530 \right] $ & $\sigma\left[\band{0.1}{r}-\band{0.1}{i}\right] = 0.07$ \cr
$I = \band{0.1}{i} - 0.1725 - 0.9718 \left[ (\band{0.1}{i}-\band{0.1}{z}) - 0.1914 \right] $ & $\sigma\left[\band{0.1}{i}-\band{0.1}{z}\right] = 0.13$ \cr
\enddata
\tablecomments{Uses AB magnitudes throughout.}
\end{deluxetable}

\begin{deluxetable}{llcl}
\tablewidth{0pt}
\tablecolumns{4}
\tablecaption{\label{derived} Header and Data Units (HDUs) of {\tt k\_nmf\_derived.default.fits}}
\tablehead{ Number & Name & Dimensions & Description}
\startdata
0 & {\tt templates} & $N_{\mathrm{basis}} \times N_t$ & Coefficients
in basis space for each template \cr 1 & {\tt spec} &
$N_{\mathrm{spec}} \times N_t$ & Smoothed spectrum of each template
\cr 2 & {\tt spec\_nl} & $N_{\mathrm{spec}} \times N_t$ & Smoothed
spectrum of each template, lines removed \cr 3 & {\tt spec\_nd} &
$N_{\mathrm{spec}} \times N_t$ & Smoothed spectrum of each template,
without dust \cr 4 & {\tt spec\_nl\_nd} & $N_{\mathrm{spec}} \times
N_t$ & Smoothed spectrum of each template, without lines or dust \cr 5
& {\tt rawspec} & $N_{\mathrm{spec}} \times N_t$ & Unsmoothed spectrum
of each template \cr 6 & {\tt rawspec\_nl} & $N_{\mathrm{spec}} \times
N_t$ & Unsmoothed spectrum of each template, lines removed \cr 7 &
{\tt rawspec\_nd} & $N_{\mathrm{spec}} \times N_t$ & Unsmoothed
spectrum of each template, without dust \cr 8 & {\tt rawspec\_nl\_nd}
& $N_{\mathrm{spec}} \times N_t$ & Unsmoothed spectrum of each
template, without lines or dust \cr 9 & {\tt lspec} &
$N_{\mathrm{spec}} \times N_t$ & emission line spectrum \cr 10 & {\tt
extinction} & $N_{\mathrm{spec}} \times N_t$ & extinction as a
function of wavelength \cr 11 & {\tt lambda} & $N_{\mathrm{spec}}$ &
Wavelength grid (\AA) \cr 12 & {\tt sfr} & $N_{\mathrm{age}} \times
N_t$ & Star-formation rate ($M_\odot$ per year) as a function of time
\cr 13 & {\tt metallicity} & $N_{\mathrm{age}} \times N_t$ & Average
metallicity as a function of time \cr 14 & {\tt ages} &
$N_{\mathrm{age}}$ & Age grid (yrs) \cr 15 & {\tt dage} &
$N_{\mathrm{age}}$ & Age bin size (yrs); {\tt sfr}$\times${\tt dage}
$=$ total star-formation in each bin \cr 16 & {\tt mass} & $N_t$ &
total mass formed in each template \cr 17 & {\tt mremain} & $N_t$ &
total current stellar mass in each template \cr 18 & {\tt mets} &
$N_t$ & average metallicity of current stars in each template \cr 19 &
{\tt m300} & $N_t$ & total stellar mass formed in last 300 Myrs in
each template \cr 20 & {\tt m1000} & $N_t$ & total stellar mass formed
in last 1 Gyr in each template \cr 21 & {\tt basis\_ages} &
$N_{\mathrm{age}} \times N_{\mathrm{mets}} \times N_{\mathrm{dust}}$ &
ages of each stellar population basis vector \cr 22 & {\tt
basis\_mets} & $N_{\mathrm{age}} \times N_{\mathrm{mets}} \times
N_{\mathrm{dust}}$ & metallicity of each stellar population basis
vector \cr 23 & {\tt basis\_dusts} & $N_{\mathrm{age}} \times
N_{\mathrm{mets}} \times N_{\mathrm{dust}}$ & dust properties of each
stellar population basis vector \cr 24 & {\tt basis\_mremain} &
$N_{\mathrm{age}} \times N_{\mathrm{mets}} \times N_{\mathrm{dust}}$ &
fraction of original stellar mass surviving for each basis vector \cr
\enddata 
\tablecomments{All quantities are in floating point (four byte)
precision, except {\tt dusts} which is an array of structures, whose
format is described in text.}
\end{deluxetable}

\clearpage
\clearpage

\setcounter{thefigs}{0}

\clearpage
\stepcounter{thefigs}

\clearpage
\stepcounter{thefigs}

\clearpage
\stepcounter{thefigs}

\clearpage
\stepcounter{thefigs}

\clearpage
\stepcounter{thefigs}

\clearpage
\stepcounter{thefigs}

\clearpage
\stepcounter{thefigs}

\clearpage
\stepcounter{thefigs}

\clearpage
\stepcounter{thefigs}

\clearpage
\stepcounter{thefigs}

\clearpage
\stepcounter{thefigs}

\clearpage
\stepcounter{thefigs}

\clearpage
\stepcounter{thefigs}

\clearpage
\stepcounter{thefigs}

\clearpage
\stepcounter{thefigs}

\clearpage
\stepcounter{thefigs}

\clearpage
\stepcounter{thefigs}

\clearpage
\stepcounter{thefigs}

\clearpage
\stepcounter{thefigs}

\clearpage
\stepcounter{thefigs}


\begin{thebibliography}{39}
\expandafter\ifx\csname natexlab\endcsname\relax\def\natexlab#1{#1}\fi

\bibitem[{{Bell} \& {de~Jong}(2001)}]{bell01b}
{Bell}, E.~F. \& {de~Jong}, R.~S. 2001, \apj, 550, 212

\bibitem[{{Bertin} \& {Arnouts}(1996)}]{bertin96a}
{Bertin}, E. \& {Arnouts}, S. 1996, \aaps, 117, 393

\bibitem[{{Bessell}(1990)}]{bessell90a}
{Bessell}, M.~S. 1990, \pasp, 102, 1181

\bibitem[{Blanton {et~al.}(2003)Blanton, Brinkmann, Csabai, Doi, Eisenstein,
  Fukugita, Gunn, Hogg, \& Schlegel}]{blanton03b}
Blanton, M.~R., Brinkmann, J., Csabai, I., Doi, M., Eisenstein, D.~J.,
  Fukugita, M., Gunn, J.~E., Hogg, D.~W., \& Schlegel, D.~J. 2003, \aj, 125,
  2348

\bibitem[{{Blanton} {et~al.}(2003){Blanton}, {Lin}, {Lupton}, {Maley}, {Young},
  {Zehavi}, \& {Loveday}}]{blanton03a}
{Blanton}, M.~R., {Lin}, H., {Lupton}, R.~H., {Maley}, F.~M., {Young}, N.,
  {Zehavi}, I., \& {Loveday}, J. 2003, \aj, 125, 2276

\bibitem[{{Blanton} {et~al.}(2005)}]{blanton05a}
{Blanton}, M.~R. {et~al.} 2005, \aj, 129, 2562

\bibitem[{Bruzual \& Charlot(2003)}]{bruzual03a}
Bruzual, A.~G. \& Charlot, S. 2003, \mnras, 344, 1000

\bibitem[{{Chabrier}(2003)}]{chabrier03a}
{Chabrier}, G. 2003, \pasp, 115, 763

\bibitem[{{Cohen} {et~al.}(2003){Cohen}, {Wheaton}, \& {Megeath}}]{cohen03a}
{Cohen}, M., {Wheaton}, W.~A., \& {Megeath}, S.~T. 2003, \aj, 126, 1090

\bibitem[{{Connolly} {et~al.}(1995){Connolly}, {Szalay}, {Bershady}, {Kinney},
  \& {Calzetti}}]{connolly95b}
{Connolly}, A.~J., {Szalay}, A.~S., {Bershady}, M.~A., {Kinney}, A.~L., \&
  {Calzetti}, D. 1995, \aj, 110, 1071

\bibitem[{{Davis} {et~al.}(2003){Davis}, {Faber}, {Newman}, {Phillips},
  {Ellis}, {Steidel}, {Conselice}, {Coil}, {Finkbeiner}, {Koo}, {Guhathakurta},
  {Weiner}, {Schiavon}, {Willmer}, {Kaiser}, {Luppino}, {Wirth}, {Connolly},
  {Eisenhardt}, {Cooper}, \& {Gerke}}]{davis03a}
{Davis}, M., {Faber}, S.~M., {Newman}, J., {Phillips}, A.~C., {Ellis}, R.~S.,
  {Steidel}, C.~C., {Conselice}, C., {Coil}, A.~L., {Finkbeiner}, D.~P., {Koo},
  D.~C., {Guhathakurta}, P., {Weiner}, B., {Schiavon}, R., {Willmer}, C.,
  {Kaiser}, N., {Luppino}, G.~A., {Wirth}, G., {Connolly}, A., {Eisenhardt},
  P., {Cooper}, M., \& {Gerke}, B. 2003, in Discoveries and Research Prospects
  from 6- to 10-Meter-Class Telescopes II. Edited by Guhathakurta, Puragra.
  Proceedings of the SPIE, Volume 4834, pp. 161-172 (2003)., 161--172

\bibitem[{Eisenstein {et~al.}(2001)}]{eisenstein01a}
Eisenstein, D.~J. {et~al.} 2001, \aj, 122, 2267

\bibitem[{{Faber} {et~al.}(2003){Faber}, {Phillips}, {Kibrick}, {Alcott},
  {Allen}, {Burrous}, {Cantrall}, {Clarke}, {Coil}, {Cowley}, {Davis}, {Deich},
  {Dietsch}, {Gilmore}, {Harper}, {Hilyard}, {Lewis}, {McVeigh}, {Newman},
  {Osborne}, {Schiavon}, {Stover}, {Tucker}, {Wallace}, {Wei}, {Wirth}, \&
  {Wright}}]{faber03a}
{Faber}, S.~M., {Phillips}, A.~C., {Kibrick}, R.~I., {Alcott}, B., {Allen},
  S.~L., {Burrous}, J., {Cantrall}, T., {Clarke}, D., {Coil}, A.~L., {Cowley},
  D.~J., {Davis}, M., {Deich}, W.~T.~S., {Dietsch}, K., {Gilmore}, D.~K.,
  {Harper}, C.~A., {Hilyard}, D.~F., {Lewis}, J.~P., {McVeigh}, M., {Newman},
  J., {Osborne}, J., {Schiavon}, R., {Stover}, R.~J., {Tucker}, D., {Wallace},
  V., {Wei}, M., {Wirth}, G., \& {Wright}, C.~A. 2003, in Instrument Design and
  Performance for Optical/Infrared Ground-based Telescopes. Edited by Iye,
  Masanori; Moorwood, Alan F. M. Proceedings of the SPIE, Volume 4841, pp.
  1657-1669 (2003)., 1657--1669

\bibitem[{Fukugita {et~al.}(1996)Fukugita, Ichikawa, Gunn, Doi, Shimasaku, \&
  Schneider}]{fukugita96a}
Fukugita, M., Ichikawa, T., Gunn, J.~E., Doi, M., Shimasaku, K., \& Schneider,
  D.~P. 1996, \aj, 111, 1748

\bibitem[{{Giavalisco} {et~al.}(2004)}]{giavalisco04a}
{Giavalisco}, M. {et~al.} 2004, \apjl, 600, L93

\bibitem[{{Hayes}(1985)}]{hayes85a}
{Hayes}, D.~S. 1985, in IAU Symp. 111: Calibration of Fundamental Stellar
  Quantities, Vol. 111, 225--249

\bibitem[{{Hogg}(1999)}]{hogg99a}
{Hogg}, D.~W. 1999, in ASP Conf. Ser. 193: The Hy-Redshift Universe: Galaxy
  Formation and Evolution at High Redshift, 224--+

\bibitem[{{Hogg} {et~al.}(2001){Hogg}, {Finkbeiner}, {Schlegel}, \&
  {Gunn}}]{hogg01a}
{Hogg}, D.~W., {Finkbeiner}, D.~P., {Schlegel}, D.~J., \& {Gunn}, J.~E. 2001,
  \aj, 122, 2129

\bibitem[{Hogg {et~al.}(2002)}]{hogg02a}
Hogg, D.~W. {et~al.} 2002, \aj, 124, 646

\bibitem[{{Jarrett} {et~al.}(2000){Jarrett}, {Chester}, {Cutri}, {Schneider},
  {Skrutskie}, \& {Huchra}}]{jarrett00a}
{Jarrett}, T.~H., {Chester}, T., {Cutri}, R., {Schneider}, S., {Skrutskie}, M.,
  \& {Huchra}, J.~P. 2000, \aj, 119, 2498

\bibitem[{{Kauffmann} {et~al.}(2003)}]{kauffmann03a}
{Kauffmann}, G. {et~al.} 2003, \mnras, 341, 33

\bibitem[{{Kennicutt} {et~al.}(1994){Kennicutt}, {Tamblyn}, \&
  {Congdon}}]{kennicutt94a}
{Kennicutt}, R.~C., {Tamblyn}, P., \& {Congdon}, C.~E. 1994, \apj, 435, 22

\bibitem[{{Kewley} {et~al.}(2001){Kewley}, {Dopita}, {Sutherland}, {Heisler},
  \& {Trevena}}]{kewley01a}
{Kewley}, L.~J., {Dopita}, M.~A., {Sutherland}, R.~S., {Heisler}, C.~A., \&
  {Trevena}, J. 2001, \apj, 556, 121

\bibitem[{Kurucz(1991)}]{kurucz91a}
Kurucz, R.~L. 1991, in Precision Photometry:\ Astrophysics of the Galaxy, L.
  Davis Press, Inc., 27--44

\bibitem[{Lee \& Seung(1999)}]{lee99a}
Lee, D.~D. \& Seung, H.~S. 1999, Nature, 401, 788

\bibitem[{Lee \& Seung(2000)}]{lee00a}
Lee, D.~D. \& Seung, H.~S. 2000, in {NIPS}, 556--562

\bibitem[{Lupton {et~al.}(2001)Lupton, Gunn, {Ivezi{\'c}}, Knapp, Kent, \&
  Yasuda}]{lupton01a}
Lupton, R.~H., Gunn, J.~E., {Ivezi{\'c}}, Z., Knapp, G.~R., Kent, S., \&
  Yasuda, N. 2001, in ASP Conf.\ Ser.\ 238:\ Astronomical Data Analysis
  Software and Systems X, Vol.~10, 269

\bibitem[{Martin {et~al.}(2005)}]{martin05a}
Martin, D.~C. {et~al.} 2005, \apjl, 619, L1

\bibitem[{{Oke} \& {Sandage}(1968)}]{oke68a}
{Oke}, J.~B. \& {Sandage}, A. 1968, \apj, 154, 21

\bibitem[{Pier {et~al.}(2003)Pier, Munn, Hindsley, Hennessy, Kent, Lupton, \&
  {Ivezi{\' c}}}]{pier03a}
Pier, J.~R., Munn, J.~A., Hindsley, R.~B., Hennessy, G.~S., Kent, S.~M.,
  Lupton, R.~H., \& {Ivezi{\' c}}, {\v Z}. 2003, \aj, 125, 1559

\bibitem[{Richards {et~al.}(2002)}]{richards02a}
Richards, G. {et~al.} 2002, \aj, 123, 2945

\bibitem[{{Schneider} {et~al.}(1983){Schneider}, {Gunn}, \&
  {Hoessel}}]{schneider83a}
{Schneider}, D.~P., {Gunn}, J.~E., \& {Hoessel}, J.~G. 1983, \apj, 264, 337

\bibitem[{Sha {et~al.}(2002)Sha, Saul, \& Lee}]{sha02a}
Sha, F., Saul, L., \& Lee, D. 2002, University of Pennsylvania, Technical
  Report MS-CIS-02-19

\bibitem[{Skrutskie {et~al.}(1997)}]{skrutskie97a}
Skrutskie, M.~F. {et~al.} 1997, in ASSL Vol. 210: The Impact of Large Scale
  Near-IR Sky Surveys, 25

\bibitem[{Smith {et~al.}(2002)Smith, Tucker, {et~al.}}]{smith02a}
Smith, J.~A., Tucker, D.~L., {et~al.} 2002, \aj, 123, 2121

\bibitem[{Stoughton {et~al.}(2002)}]{stoughton02a}
Stoughton, C. {et~al.} 2002, \aj, 123, 485

\bibitem[{Strauss {et~al.}(2002)}]{strauss02a}
Strauss, M.~A. {et~al.} 2002, \aj, 124, 1810

\bibitem[{{Witt} \& {Gordon}(2000)}]{witt00a}
{Witt}, A.~N. \& {Gordon}, K.~D. 2000, \apj, 528, 799

\bibitem[{York {et~al.}(2000)}]{york00a}
York, D. {et~al.} 2000, \aj, 120, 1579

\end{thebibliography}
\end{document}